\documentstyle[manuscript,eqsecnum,aps,version2,epsf]{revtex}

\begin{document}
\draft
\preprint{OU-HEP STB1-96}
\title
\bf Ultrametricity and Memory\\  
in a  Solvable Model of Self-Organized Criticality
\endtitle

\author{Stefan Boettcher$^{1,2}$ and Maya Paczuski$^1$}
\instit
$^1$Department of Physics, Brookhaven National Laboratory,
Upton, NY 11973\\
$^2$Department of Physics and Astronomy, The
University of Oklahoma, Norman, OK 73019-0225
\endinstit
\medskip
\centerline{\today}
\medskip

\abstract 
Slowly driven dissipative systems may evolve to a critical
state where long periods of apparent equilibrium are punctuated by
intermittent avalanches of activity.  We present a self-organized
critical model of punctuated equilibrium behavior in the context of
biological evolution, and solve it in the limit that the number of
independent traits for each species diverges.  We derive an exact
equation of motion for the avalanche dynamics from the microscopic
rules.  In the continuum limit, avalanches propagate via a diffusion
equation with a nonlocal, history-dependent potential representing
memory.  This nonlocal potential gives rise to a non-Gaussian (fat)
tail for the subdiffusive spreading of activity.  The probability for
the activity to spread beyond a distance $r$ in time $s$ decays as
$\sqrt{24\over\pi}s^{-3/2}x^{1/3} \, \exp{[-{3\over 4}x^{1/3}]}$ for
$x={r^4\over s} \gg 1$.  The potential represents a
hierarchy of time scales that is dynamically
generated by the ultrametric structure of avalanches, which can be
quantified in terms of ``backward'' avalanches.  In addition, a number
of other correlation functions characterizing the punctuated
equilibrium dynamics are determined exactly.  \endabstract




\section{Introduction}
Many natural phenomena evolve intermittently rather than following a
uniform, gradual path.  In particular the dynamics out of
equilibrium may follow a
step-like pattern with long, dormant plateaus interrupted by sudden
bursts, or avalanches, where the accumulated stress is released.
Avalanche dynamics violates the picture of gradualism where large
systems evolve continuously, for instance to a local energy minimum.
The bursts which separate subsequent metastable states
may eventually span all scales up to the system size \cite{soc}
thus providing a general mechanism for  long
range spatiotemporal correlations.  The
appearance of fractals, $1/f$ noise, Levy flights, etc., have been
unified by relating these phenomena, in a broad class of
non-equilibrium models, to an underlying avalanche structure
\cite{scaling}.

Even though the theory of uniformitarianism, or gradualism, has 
historically dominated both geology and paleontology, prototypical 
examples of avalanche dynamics lie in these two domains
\cite{complexity}.  For instance, the distribution of earthquake
magnitudes follows a  power law found by Gutenberg and Richter
\cite{G+R}.  The scale-free  variation from small events to large events
indicates a common  dynamical origin and eventually led to the
suggestion that  earthquakes are an example of self-organized
criticality \cite{earthquake}.  Most of the time, the crust of the earth
appears stable. These periods of  apparent equilibrium are punctuated by
earthquakes, which take  place on a fractal fault structure 
\cite{barton} that stores information about the history of the system.

Actually, over twenty years ago, Gould and Eldredge proposed that
biological evolution also takes place in terms of punctuations, where
many species become extinct and new species emerge, interrupting
periods of low activity \cite{Gould}.  In this context, punctuated
equilibrium usually refers to the intermittent dynamics of single species,
where morphological change is concentrated in short intervals in time
interrupting long periods of stasis.  These punctuations may be
correlated to large extinction events in the global ecology, which may
themselves be distributed according to a power law analogous to the
Gutenburg-Richter law for earthquakes \cite{complexity,sfbj}.

This view was promoted by Bak and Sneppen \cite{BS} who introduced a
simple self-organized critical (SOC) model for coevolutionary avalanches of
different species in an ecology.  The model explicitly treats
macroevolution as a many-body statistical problem where the fitness
landscape in which each species evolves is affected by changes in
other species in the ecology.  The mutation of the ``least fit''
species in the system completely changes the fitness landscape of some other
species and coevolution with punctuated equilibrium behavior is
obtained.  This occurs without fine tuning parameters and without the
need for external shocks leading to cataclysms of mass extinction.
The extinction events (avalanches) have a power law distribution of
sizes, where most extinction is concentrated in the largest
events.  This provides some theoretical underpinning for catastrophism
rather than uniformitarianism as the defining characterization of
evolutionary history \cite{uniform}.  The model may capture some
robust features of real evolution, such as punctuated equilibrium
and catastrophism, in spite of its drastic simplifications.

 In fact Ito \cite{ito} has related the spatiotemporal pattern of 
earthquakes in California to the avalanche pattern found in the  
Bak-Sneppen evolution model.  The data for the pattern of successive
earthquakes over time may be  fractal in space and time, so that each
earthquake can be viewed a single event within a much larger avalanche
structure consisting of many earthquakes.  If this picture is correct,
earthquakes can be described as a ``fractal renewal process''
\cite{scaling,lowen} with a power law 
distribution of times between subsequent earthquakes in a given
region. The analogy can be made by replacing the least fit species with
the weakest link in the earth's crust. 
Ito's observations of return times for earthquakes in California
 are consistent with general  scaling relations valid for
the Bak-Sneppen model, as well as other  extremal SOC models \cite{scaling}
 for invasion percolation \cite{ip}, interface depinning
\cite{sneppen}, and flux creep \cite{zaitsev}.

Although the Bak-Sneppen model is an extremely simple model of SOC
and a few exact results as well as many scaling relations are known
for it \cite{scaling}, it has not (yet) been solved.
A variety of similar evolution models have been subsequently introduced to
incorporate more aspects of reality such as speciation and external shocks
\cite{otherbsmodels}. These models have been primarily studied numerically.

Recently, we have proposed a multi-trait evolution model with $M$
independent, internal degrees of freedom for each species \cite{BoPa}.
This model is solvable in the limit $M\rightarrow\infty$; it includes
the Bak-Sneppen model when $M=1$.  We have previously reported some of
the simplest results \cite{BoPa}.  Here we present a more complete
analysis of the model.  Note that the $M\rightarrow\infty$ limit is
not mean-field theory \cite{meanfield} because it contains spatial
correlations and punctuated equilibrium dynamics.  In addition, we
find rigorous results that might be too delicate to detect with
numerical calculations alone, such as the non-Gaussian tail for the
distribution of activity.

In the multi-trait evolution model, we have incorporated the notion
that the survivability of each species depends on a number of
independent traits associated with the different tasks that it has to
perform in order to survive \cite{Kauffman}.  Evolution proceeds via
an extremal dynamics where the species in the global ecology with the
lowest barrier to mutation, or the least fit, mutates.  This event
affects certain barriers to mutation or fitnesses of other species in
the system which are related through, for instance, a food chain.  As
a consequence of the interaction between species, even species that
possess well-adapted abilities, with high barriers, can be undermined
in their existence by weak species with which they interact.  This may
lead to a chain reaction of coevolution.  The pattern of change of
individual species exhibits punctuated equilibrium behavior which
comes from episodes of mass extinction events sweeping through the
species.  Punctuated equilibrium is described by a Devil's staircase, as
shown in Fig.~\ref{devil}.  The introduction of many internal traits
for each species is consistent with paleontological observations
indicating that evolution within a species is ``directed''
\cite{Kauffman}; morphological change over time is concentrated in a
few traits, while most others traits of the species are static.

Our main analytical results are as follows: For $M\rightarrow \infty$,
we derive an exact equation of motion, given in Eq.~(\ref{envelope}),
for the macroscopic observables in the SOC state from the microscopic
dynamics.  From this equation, which is our central result,
one can extract separate equations for
the temporal and spatial distribution of avalanche sizes, both of which 
are determined exactly in Eqs.~(\ref{z3},\ref{z4}) and were previously derived
by us in Ref.~\cite{BoPa}.  In the continuum limit, the subdiffusive
dynamics is given by a ``Schr\"odinger'' equation 
with a nonlocal potential in time, Eq.~(\ref{nonlocaleq}).  This potential
represents memory. The exact
asymptotic solution, Eq.~(\ref{nongaussian}), of the Schr\"odinger
equation at large length and time scales has a non-Gaussian tail.
This solution describes the nontrivial spatiotemporal pattern of
activity in the self-organized critical state. The anomalous
diffusion arises from a long-term memory effect due to the
ultrametric tree structure \cite{ultram} of avalanches.  
This tree structure of activity is quantified by calculating the
ultrametric distances between subsequent extremal (minimal) sites.
The probability distribution for this distance is a power law at large
times and asymptotically approaches the probability distribution of
all backward avalanches.
This latter quantity, which we will define in the text, is calculated exactly in
Eq.~(\ref{allbackward}).  A number of other distribution functions
are also determined.  The critical exponents are $D=4$, $\tau=3/2$,
$\tau_R = 3$, $\gamma=1$, $\nu=\sigma=1/2$, $\tau_f^{all}=2$,
$\tau_b^{all}=3/2$, and $\tau_{FIRST}=2-d/4$.  The nonlocal time
dependence of the dynamical equations 
and the ultrametric structure of avalanches suggest a possible
relation between glassy dynamics and self-organized criticality
\cite{preprint}.

In Section II, we introduce our model and show that it self-organizes
to the critical state.  It is demonstrated that avalanches in the
critical state have an ultrametric tree structure. In Section III, we
derive the equation of motion for the critical state for the
$M\rightarrow\infty$ model and present our main analytical results for
the anomalous diffusion.  In Section IV, we show that the
critical behavior is characterized by simple power laws with
specific exponents that verify general scaling relations for
nonequilibrium phenomena.  In Section V we calculate the distributions
for ``backward avalanches.''  Due to irreversibility these differ from
the usual ``forward'' avalanche distributions.  The backward
avalanches can be related to the ultrametric distances between
subsequent activity.  Most details of our calculations have
been deferred to the Appendices.

\section{The Multi-Trait Evolution Model}

Our model is defined as follows: A species is represented by a single
site on a $d$-dimensional lattice.  The collection of traits for each
species is represented by a set of $M$ numbers in the unit interval. A
larger number represents a better ability to perform that particular
task, while smaller numbers pose less of a barrier against mutation.
Therefore, we ``mutate'' at every time step the smallest number in the
entire system by replacing it by a new (possibly smaller) number that is
randomly drawn from a flat distribution in the  unit interval, ${\cal
P}$. Choosing the smallest random number mimics the Darwinian principle
that the least fit species mutates \cite{Darwin}.  

The dynamical impact of this event on neighboring species is simulated
by also replacing one of the $M$ numbers on each neighboring site with
a new random number drawn from ${\cal P}$.  Which one of the $M$ numbers
is selected for such an update is determined at random, since we assume
that a mutation in the traits of one species can lead to an adaptive
change in any one of the traits of a dependent species.  The interaction
between the fitnesses of species leads to a chain reaction of coevolution.
  

\subsection{Self-organization}

The sequence of selective random updates at extremal (minimal) sites
with nearest-neighbor interactions drives the system from any initial
state to a self-organized critical state  in which species 
exist in a state of punctuated equilibrium with bursts of evolutionary
activity that are correlated over all spatial and temporal extents. In
this state almost all species have reached fitnesses above a SOC
threshold, enjoying long periods of quiescence, interrupted by
intermittent activity when changes in neighboring species force a
readjustment in their own barriers. A snapshot of the stationary state
in a finite one dimensional system is shown in Fig. \ref{stationary}. 

The self-organization process for the finite $M$ model is similar to
that for the Bak-Sneppen model ($M=1$). It is described by a ``gap''
equation that relates the rate of approach towards the stationary
attractor to the average avalanche size.  This equation demonstrates
that the stationary state of the system is a critical state for the
avalanches, where the average avalanche size diverges. Following
Ref.~\cite{scaling} we define the gap, $G(s)$, to be the largest of the
minimal random numbers selected up to time $s$ given an initial state at
$s=0$ where all the random numbers are uniformly distributed in the unit
interval.  Therefore, $G(s=0)= {\cal O}(1/ML^d)$.  As evolution
proceeds, the gap $G(s)$ increases monotonically in a stepwise fashion
with intermediate plateaus that become longer and longer.  These 
plateaus occur when subsequent minimal random numbers in the system  are
smaller than a minimum at some previous time. One can assign to  each
plateau an avalanche which starts and ends with consecutive changes in
$G(s)$, and which consists of all the random numbers in the gap below
$G(s)$.  As the gap increases, the probability for the new random
numbers to fall below the current gap increases also, and longer and
longer avalanches typically occur.

Following Ref.~\cite{scaling} again, it can be shown from the law of large 
numbers that in the limit of large system sizes $L$, the growth of the 
gap versus time $s$ obeys  the   gap equation:
\begin{eqnarray}
{d G(s) \over d s } = {1- G(s) \over M L^d <S>_{G(s)}} \qquad .
\label{e.approach}
\end{eqnarray}
As the gap increases, so does the average avalanche size $<S>_{G(s)}$,
which eventually diverges as $G(s) \rightarrow G_c$. In the limit $L
\rightarrow \infty$, the density of sites with random numbers less
than $G_c$ vanishes, and the distribution of random numbers is uniform
above $G_c$. The gap equation (\ref{e.approach}) contains the
essential physics of SOC phenomena.  When the average avalanche size
diverges, $<S>_{G(s)} \rightarrow \infty$, the system becomes
critical. At the same time, $d G \over d s$ approaches zero, which
means that the system becomes stationary. For any finite $M$ the
process of self-organization is the same as for the $M=1$ model, and
all of the results derived for that case apply. Since $M$ just enters
as a rescaling of the system size $L$ in Eq.  (\ref{e.approach}), it
is plausible that the critical behavior for any finite $M$ is in the
same universality class as for $M=1$.  We have shown \cite{BoPa} that
the limit $M \rightarrow \infty$ is a different universality class.

\subsection{Ultrametricity of Avalanches}

It is useful to consider the case where, at a certain time, the smallest 
random number in the system has the value $\lambda$. A $\lambda$ 
avalanche by definition consists of all subsequent random numbers which
are below $\lambda$. The $\lambda$ avalanche that started at $s$ ends
at the first instant $s+s'$ when the smallest random number in the
system is larger than $\lambda$.  All of the random numbers that are
below the threshold value $\lambda$ at any one instant in time are
called ``active'' because they make up the $\lambda$ avalanche. 

We now consider the sequence of minimal values $\lambda_{min}(s)$
comprising any $\lambda$ avalanche.  Each value $\lambda_{min}(s)$ has
a parent barrier value $\lambda_{min}(s-s')$ preceding it within the
$\lambda$ avalanche.  This parent is the barrier that introduced the
particular random number into the system that became the minimum at
time $s$.  Obviously, the parent of the $\lambda_{min}(s)$ value also
has its own parent.  One can therefore place all of the barrier values
within a given avalanche onto a tree, as shown in
Fig.~\ref{ultratree}.  The distance on the tree between any two active
barrier values at a given time is determined by the most recent common
ancestor of the two values.  This distance is ultrametric
\cite{ultram}.  In Section V
we relate the probability distribution of the ultrametric distances
between subsequent minimal random numbers to the distribution of all
backward avalanches, which is determined exactly.

{}For finite $M$, active barriers can be eliminated both by becoming the
global minimum at some time, or by being chosen 
in the nearest-neighbor interaction when the global minimum occurs on a
neighboring site.  The case $M\to\infty$ of the model is special because
the existing active barriers that any species possesses can only be
changed if these barriers themselves become the global minimum, despite
the nearest-neighbor interaction in the model.  Since there are
infinitely many barriers {\sl on each site}, no existing  active barrier
is ever likely to be chosen for an update in a nearest-neighbor
interaction. The nearest-neighbor interaction can only create new active
barriers. This fact allows us to formulate  a cascade mechanism which
can be solved exactly. 

In what follows we shall consider the stationary state behavior in the
limit of infinite system size, and confine our presentation to the
$d=1$ dimensional case of our model with $M\rightarrow\infty$.  We
point out later which critical exponents are dimension dependent and
dimension independent, and we will discuss in more detail the results
for higher dimensional lattices elsewhere \cite{else}.  To simplify
the algebra further, we make a slight modification of the model
without restricting the generality of the results: At each time step
during the avalanche, the smallest active barrier is set to unity
instead of being replaced by a new random number. Then, there is
either no new active barrier created with probability $(1-\lambda)^2$,
or one such barrier is created to the left or to the right of the
minimum with probability $\lambda(1- \lambda)$ in either case, or two
active barriers are created to the left and the right of the minimum
with probability $\lambda^2$.

The spatial and temporal correlations in our model can be separated into
two independent equations of motion for the width and duration of
avalanches. For instance, the distribution of avalanches sizes $s$ is
given by (see Appendix A) 
\begin{eqnarray}
P_{\lambda}(s)&=& 4^s\lambda^{s-1}(1-\lambda)^{s+1} A(s) \nonumber \\
A(s)&=& {\Gamma\left(s+{1\over 2}\right)\over
\Gamma\left(1\over 2\right)\Gamma\left(s+2\right)} \qquad . \label{z3}
\end{eqnarray}
For $\lambda=\lambda_c=1/2$, $P_{\lambda_c}(s) \sim s^{-\tau}$ for
large sizes with $\tau=3/2$.  For $\lambda <\lambda_c$, the critical
avalanches are subdivided into smaller avalanches and the distribution
acquires a cutoff.  The critical behavior is quite different than the
$M=1$ model where the numerical result is $\tau \simeq 1.1$ in one
dimension \cite{scaling}. The model for $M=\infty$ clearly represents
a different universality class than the original Bak-Sneppen model.
Actually, the temporal behavior in this case is identical to the
temporal behavior of the random-neighbor evolution model
\cite{meanfield}.

Unlike the random-neighbor model, though, the $M\to\infty$ model also
exhibits spatial correlations, leading to punctuated equilibrium
behavior. In Ref.~\cite{BoPa} we found that the probability
$f_r$ of a $\lambda_{\rm c}$ avalanche to ever affect a particular site
of distance $r$ from its origin is exactly given by 
\begin{eqnarray}
f_r={12\over (r+3)(r+4)},
\label{z4}
\end{eqnarray}
implying that the probability for an avalanche to reach precisely to a
site of distance $r$ falls asymptotically as $r^{-\tau_R}$ with
$\tau_R=3$. Noting that one particular site can not fully contain the
spread of an avalanche, we have obtained an equation for the probability 
$f(0,r)$ of a $\lambda_{\rm c}$ avalanche to start at the origin $i=0$
and to ever leave a box of radius $i=r$ around its origin. The leading
asymptotic behavior of the solution for confined avalanches is given by
\begin{eqnarray}
f(0,r)\sim {1\over 3}\left[{\Gamma({1\over 6})\Gamma({1\over 2})
\over\Gamma({2\over 3})}\right]^2{1\over r^2}\qquad (r\to\infty), 
\label{Z5}
\end{eqnarray}
confirming that $\tau_R=3$. The calculation leading to Eq.~(\ref{Z5}) is
given in Appendix B. A comparison of numerical results for $f(0,r)$ with
the asymptotic behavior, given in Fig.~\ref{f.box}, show perfect
agreement.

\section{The Cascade Equation and Main Results}

Ultimately, spatial and temporal correlations are interrelated through
the microscopic rules for the propagation of activity in space and
time.  Ideally, one would like to determine the propagator $G(r,s)$
which is the probability that the minimum barrier value will be
located at position $r$ at time $s$ given that it was at location $0$
at time $0$ for the infinite $\lambda_c$ avalanche.  We have not been
able to analytically calculate this quantity directly.  We have
instead focussed on a different quantity.  Let $F_{\lambda}(r,s)$ be
the probability for a $\lambda$ avalanche to survive precisely $s$
steps and to have affected a particular site of distance $r$ from its
origin. Conceptually, the quantity $F_{\lambda_c}(r,s)$ may roughly
correspond to an envelope function of the propagator $G(r,s)$. This
relation between $F$ and $G$ is explained in Fig.~\ref{box}. Due to
the lack of any scale in the model, it is plausible that the
asymptotic behavior of $G$ and $F$ is identical, as comparison with
numerical calculations suggests.

The direct analysis of this envelope function proves to be rewarding in 
many respects.  We find a cascade equation which can be reduced to
separate equations for spatial and temporal correlations. In the
continuum  limit, the avalanche dynamics is given in terms of
a Schr\"odinger equation with a nonlocal potential.
We solve this equation to find the 
leading  asymptotic behavior of $F_{\lambda_c}(r,s)$. This calculation
yields a non-Gaussian tail in the distribution with an avalanche
dimension  $D=4$, signaling subdiffusive behavior for the spread of
activity.

{}First, we consider $P_{\lambda}(r,s)$, the probability  that the
$\lambda$ avalanche dies precisely after $s$ updates and does  {\it not}
affect a particular site $r$ away from the origin of the avalanche. The
quantities $P_{\lambda}(r,s)$ and $F_{\lambda}(r,s)$ are simply related:

\begin{equation}
P_{\lambda}(r,s)= P_{\lambda}(r=\infty, s) - F_{\lambda}(r,s) ,
\label{deffrs}
\end{equation}
where $P_{\lambda}(r=\infty,s)=P_{\lambda}(s)$, as given in
Eq.~(\ref{z3}). Since we consider the avalanche to start with a single
active barrier at $r=0$ and $s=0$, it is $P_{\lambda}(r=0,s)\equiv 0$
for all $s\geq 0$, and $P_{\lambda}(r,s=0)\equiv 0$ for all $r$.
The remaining properties of a $\lambda$ avalanche  can be
deduced from the properties of avalanches that ensue after the first
update. It will terminate with probability $(1-\lambda)^2$ after the
first update when the update does not produce any new active barriers.
Thus, it is $P_\lambda(r,s=1)\equiv (1-\lambda)^2$. For avalanches
surviving until $s\geq 2$ we find for $r\geq 1$
\begin{eqnarray}
P_\lambda(r,s)&=&{\lambda(1-\lambda)}\left[P_\lambda(r-1,s-1)
+P_\lambda(r+1,s-1)\right]\cr
\noalign{\medskip}
&&\quad+{\lambda^2}\sum_{s'=0}^{s-1} P_\lambda(r-1,s')  P_\lambda(r+1,s-1-s')
\label{y1}
\end{eqnarray}
in the following way: The first update may create exactly one new active 
barrier with probability $\lambda(1-\lambda)$ either to the left or to
the right of the origin (i. e. one step towards or away from the chosen
site of distance $r$).  In this case, the properties of the original
avalanche of duration $s$ are related to the properties of  an avalanche
of duration $s-1$ with regard to a site of distance $r-1$  or $r+1$,
respectively. Finally, the first update may create two new  active
barriers with probability $\lambda^2$ to the left and the right  of the
origin. Then, the properties of the original avalanche of duration $s$
are related to the properties of all combinations of two avalanches of
combined duration $s-1$. Both of these avalanches evolve  in a  {\sl
statistically independent}  manner for $M=\infty$.  Since  only one of
these avalanches can be updated at each time step, their  combined
duration has to add up to $s-1$ for this process to contribute  to the
avalanche of duration $s$. For any such combination, the probability to
not affect the chosen site of distance $r$ from the origin is given
simply by the product of the probabilities for the two ensuing
avalanches to not affect a chosen site of distance $r-1$ or $r+1$,
respectively.

Before proceeding with the solution of Eq.~(\ref{y1}), we review some 
limiting cases to obtain previously derived results \cite{BoPa}.  
Considering the limit $r=\infty$, for $s\geq 2$, Eq.~(\ref{y1}) reduces to 

\begin{eqnarray}
P_{\lambda}(s)= 2\lambda(1-\lambda) P_{\lambda}(s-1)+\lambda^2
\sum_{s'=0}^{s-1} P_{\lambda}(s') P_{\lambda}(s-1-s').
\label{E1}
\end{eqnarray}
This is the cascade equation for the lifetime distribution of avalanches
whose solution is Eq. (\ref{z3}) (see Appendix A).
Near the critical point, $\Delta\lambda=\lambda_c-\lambda\to 0_+$, the lifetime
distribution obeys a scaling form 
\begin{equation}
P_{\lambda}(s) \sim s^{-\tau} G\left(s\,\Delta\lambda^{1\over\sigma}\right)
\qquad \left(\tau={3\over 2}, \sigma={1\over 2} \right).
\end{equation}

Similarly, we can rederive Eq.~(\ref{z4}) for the spatial 
correlations. Defining $N_\lambda(r)= \sum_s P_{\lambda}(r,s)$, 
Eq.~(\ref{y1}) yields $N_\lambda(0)=0$, and for $r\geq 1$
\begin{equation}
N_\lambda(r)=(1-\lambda)^2 +\lambda(1-\lambda)\left[N_\lambda(r-1) + 
N_\lambda(r+1)\right] + \lambda^2N_\lambda(r-1)N_\lambda(r+1) . 
\label{n1}
\end{equation}
$N_\lambda(r)$ is the probability for an avalanche of any temporal
extent not  to reach a particular point of distance $r$ before it dies.
Eq.~(\ref{n1}) for  $\lambda=\lambda_{\rm c}=1/2$ is solved by
$N_{\lambda_{\rm c}}(r)=1-f_r$  with $f_r$ given in Eq.~(\ref{z4}).
Close to the critical point this quantity also obeys a scaling form: 
\begin{equation}
1- N_\lambda(r) \sim {1\over r^{\tau_R -1}} H\left(r\, \Delta 
\lambda^{\nu}\right) \qquad \left(\tau_R=3, \nu={1\over 2} \right).
\label{fscaling}
\end{equation}

The equation governing the envelope function $F_{\lambda}(r,s)$ is
obtained by inserting Eq.~(\ref{deffrs}) into Eq.~(\ref{y1}). It is
$F_{\lambda}(0,s)\equiv P_{\lambda}(s)$, $F_{\lambda}(r,0)\equiv 0$,
$F_{\lambda}(r\geq 1, s=1)=0$,
and for $s\geq 1$, $r\geq 1$,
\begin{eqnarray}
F_{\lambda}(r,s+1)&=& {\lambda(1-\lambda)}\left[F_{\lambda}(r-1,s)
+F_{\lambda}(r+1,s)\right]\cr
\noalign{\medskip}
&&\quad+{\lambda^2}\sum_{s'=0}^{s} P_{\lambda}(s-s')\left[F_{\lambda}(r-1,s')
+F_{\lambda}(r+1,s')\right]\cr
\noalign{\medskip}
&&\quad-{\lambda^2}\sum_{s'=0}^{s} F_{\lambda}(r-1,s') F_{\lambda}(r+1,s-s').
\label{envelope}
\end{eqnarray}

Now we will focus on the spatiotemporal correlations at the critical
point $\lambda=\lambda_{\rm c}=1/2$. For sufficiently large values of
$r$ and $s$ we will show that $F_{\lambda}(r,s)\to 0$ for $r\to\infty$
sufficiently fast such that we can neglect the nonlinear term in
Eq.~(\ref{envelope}). We can take the continuum limit and obtain \cite{sigma}
\begin{equation} 
{\partial F(r,s)\over \partial s}\sim {1\over 2}\nabla_r^2 F(r,s) +
{1\over 2} \int_0^s V(s-s')F(r,s')ds' , 
\label{nonlocaleq} 
\end{equation} 
a ``Schr\"odinger'' equation in imaginary time, 
for $F(r,s)$ with a
nonlocal memory kernel $V(s)=P(s)-2\delta(s)$, where $P(s)=P_{\lambda_c}(s)$ 
given in Eq. (\ref{z3}), and
$\delta(s)$ is the usual Dirac $\delta$ function.  Note that it is
the statistical independence of the avalanches that gives $V(s)$ in terms
of the probability distribution of avalanche sizes.  The memory in the
system is characterized solely in terms of this distribution.

This nonlocal potential with the integral kernel $V(s)$ contains all
of the history dependence of the process.  In its absence the system
would have no memory and be purely diffusive with a Gaussian tail $F
\sim e^{-r^2/2s}$. In its presence the probability to have reached a
site at distance $r$ at time $s$ gets contributions from avalanches
that reached $r$ at earlier times $s'<s$.  These contributions are
weighted according to $P(s-s')$ which has a power law tail.
Avalanches contributing to $F(r,s)$ consist of sub-avalanches, one of
which reaches $r$ in time $s'$ while the other's combined duration is
$s-s'$.  The sub-avalanche tree structure gives a hierarchy of time
scales.  This changes the relaxation dynamics to be non-Gaussian, and
we find as our main result that
\begin{equation} 
F_{\lambda_{\rm c}}(r,s) \sim \sqrt{24\over\pi} s^{-{3\over 2}}
\left({r^4\over s}\right)^{1\over 3} e^{-{3\over 4} \left({r^4\over
s}\right)^{1\over 3}} \qquad\left(r^4 \gg s \gg 1\right) \quad . 
\label{nongaussian} 
\end{equation} 
It is immediately clear from the form of the solution that the avalanche
dimension for the subdiffusive spreading of activity $(r^D \sim s)$ is
$D=4$. Diffusion is slowed down because the activity has a tendency to
revisit sites, and the system remembers these previously visited sites. 
Considering $x=r^D/s$ as the scaling variable, the variation of the 
exponent is much slower with a ``fat'' tail ($\sim x^{1/3}$) compared to a 
Gaussian tail.

In Appendix C we derive the complete leading asymptotic behavior given in
Eq.~(\ref{nongaussian}).  Here we will just show how the history
dependence in the Schr\"odinger equation~(\ref{nonlocaleq}) gives rise
to the non-Gaussian tail in the exponential  of Eq.~(\ref{nongaussian}).

\subsection{Calculation of the Fat Tail}

Using a Laplace transform, ${\tilde F}(r,y)=\int_0^\infty ds~e^{-
ys}F(r,s)$, Eq.~(\ref{nonlocaleq}) turns into an ordinary second-order
differential equation in $r$,
\begin{equation}
\nabla_r^2 {\tilde F}(r,y)\sim\left[ 2y-{\tilde V}(y)\right] {\tilde
F}(r,y),
\end{equation}
where ${\tilde V}(y)$ is the Laplace transform of $V(s)$.
Since $F(r,s)$ is falling for large $r$, it is
\begin{equation}
{\tilde F}(r,y)\sim C(y) \exp\left[-r \sqrt{2y-{\tilde V}(y)}\right],
\end{equation}
and we assume that $C(y)$ is a sufficiently well-behaved function near $y=0$.

The inverse Laplace transform yields a contour integral with a contour
extending just to the right ($\eta>0$) of the imaginary $y$-axis:
\begin{equation}
F(r,s)\sim\int_{-i\infty+\eta}^{i\infty+\eta}{dy\over 2\pi i} C(y)
\exp\left[ys-r \sqrt{2y-{\tilde V}(y)}\right].
\end{equation}
The limit of large $s$ corresponds to small values of $y$ where ${\tilde
V}(y)\sim -2\sqrt{y}$ such that $\sqrt{2y-{\tilde V}(y)}\sim \sqrt{2}y^{1/4}$.
Note that the contribution from the left-hand side of Eq.~(\ref{nonlocaleq})
does not effect the leading order which merely consists of a balance between
the terms on the right-hand side. After rescaling $y\to y/s$ it is
\begin{equation}
F(r,s)\sim {C(0)\over 2\pi is}\int_{\cal C} dy~\exp\left[
y-\sqrt{2}\left({r\over s^{1\over 4}}\right) y^{1\over 4}\right],
\label{saddle}
\end{equation}
where $\cal C$ is a small piece of contour that crosses the real 
$y$-axis from below just to the right of $y=0$. It emerges that $F(r,s)$ is
exponentially cut off when $r\gg s^{1/4}\gg 1$, determining that the
avalanche dimension is $D=4$. In that limit, we can perform a steepest-descent 
analysis of the integral \cite{BeOr}. First, we note that the
expression in the exponent has a moving  saddle point at $y_0={1\over
4}r^{4/3}s^{-1/3}$. To fix the saddle point, we substitute $y=y_0 v$ to
obtain
\begin{eqnarray}
F(r,s)\sim {C(0)y_0\over 2\pi is}\int_{\cal C} dv \exp\left[y_0(v-
4v^{1\over 4})\right],
\end{eqnarray}
where the contour $\cal C$ is deformed such that it crosses the real
axis on the saddle point at $v_0=1$. To find the steepest-descent path
for the contour in the neighborhood of the saddle point, we set $v\sim
1+\epsilon+i\delta$ with $\epsilon,\delta\ll 1$. Thus, in the exponent,
it is $v-4v^{1/4}\sim \left(-3+{3\over 8}\epsilon^2-{3\over
8}\delta^2\right)-i\left({3\over 4}\epsilon\delta\right)$, indicating
that the steepest-descent path (which is always also a constant-phase
path) is given by $\epsilon\equiv 0$ in the neighborhood of the saddle
point. Thus, we substitute $v=1+i\delta$ in that neighborhood and obtain the
non-Gaussian tail in Eq.~(\ref{nongaussian}),
\begin{eqnarray}
F(r,s)\sim C'(r,s) \exp\left[-{3\over 4}\left({r^4\over
s}\right)^{1\over 3}\right] \quad (r^4\gg s\gg 1), 
\label{n4}
\end{eqnarray}
where $C'(r,s)$ only contains powers of $r$ and $s$. The function $C'(r,s)$,
as well as the behavior of $F(r,s)$ for $s\gg r^4\gg 1$, is determined in
Appendix C. In Appendix D we will also  consider the
distribution for avalanches that are fully confined in a box of size $r$
(see Fig.~\ref{box}).  We find that the dominant relaxation behavior 
is also  given by Eq. (\ref{n4}).

\subsection{Numerical comparison with the propagator}
We are now in a position to attempt to compare these results
with numerical measurements of the actual propagator $G(r,s)$. Based on earlier
numerical investigations \cite{letterfig}, we assume  that $G(r,s)$ has
the scaling form
\begin{equation}
G(r,s)\sim r^\delta g\left({r^4\over s}\right)\quad 
\left(r^4/s\gg 1\right). 
\label{scal}
\end{equation}
At each update $s$ we determine the location $r$ of the minimal random
number in a surviving avalanche that started at $s=0,~r=0$. We bin
counts as a function of $r$ and $r^4/s$ for avalanches of duration
$s<10^8$. In Fig.~\ref{scaling} we plot the numerical results for
$G(r,s)$ for various values of $r^4/s$ as a function of $s$.  This makes
the numerically imposed cut-off at $s=10^8$ apparent. The data indicates
that $G(r,s)$ rises linearly with $r$ for each value of $r^4/s$;
thus $\delta=1$. In Fig.~\ref{collapse} we plot $G(r,s)/r$ {\it vs.} $s$
for different values of $r^4/s$ for $10^4<s<5\times 10^7$. Note that all 
data ``collapses'' onto plateaus whose mean value and standard deviation
is given by the crosses with error bars to the right of each curve. 

We attempted to numerically determine the asymptotic tail of the function $g$ 
for large values of $x=r^4/s$ by assuming in accordance with 
Eq.~(\ref{nongaussian}) that 
\begin{eqnarray}
g(x)\sim e^{-A x^\alpha},\quad x\gg 1.
\label{gx}
\end{eqnarray}
{}From the sequence of plateau values in Fig.~\ref{collapse}
for $g(x)$ we determine a sequence of extrapolants
\begin{eqnarray}
{\ln\left(-\ln{g(x)}\right)\over \ln x}\sim \alpha +{\ln
A\over\ln x},\quad x\gg 1.
\label{extraseq}
\end{eqnarray}
In Fig.~\ref{extrapolation} we plot this sequence of extrapolants as a
function of $1/\ln x$ and estimate that $\alpha=0.35\pm 0.03$, in reasonable
agreement with with the value $\alpha=1/3$ from the exponential fall-off for
$F(r,s)$. Thus, the behavior of $G(r,s)$ is consistent with the non-Gaussian 
asymptotic behavior of the envelope function $F(r,s)$.

\section{Scaling Relations}

In the SOC state, spatial and temporal correlations are interrelated.
This interrelation is expressed via scaling
relations. In a broad class of  SOC models, including the
evolution models, the knowledge of just two  scaling coefficients, such
as $\tau$ and $D$, is sufficient to determine any other known
coefficient of the SOC state,  including the approach to the attractor,
through these scaling  relations \cite{scaling}.  For example, we have
shown in Sec. III  that the activity in the SOC state spreads in a
subdiffusive manner,  $r \sim s^{1/D}$, where $D=4$ is the avalanche
dimension. In Ref.~\cite{BoPa}  we have numerically determined the 
root-mean-square distance of the location  of the activity at time $s$ and
shown that it scales as $s^{1/4}$. In  Fig.~\ref{compact} we show that
 the number of sites covered by the  activity also grows as
$s^{1/4}$. This verifies that rather than being multifractal
there is only one  dimension for the
avalanche.

The probability distribution of $\lambda_c$ avalanche duration  is a power law 
with exponent $\tau=3/2$, and we have shown that the
probability distribution  of spatial extents of avalanches is also
a power law with critical  exponent $\tau_R=3$.  This verifies the
scaling relation $\tau_R -1 =  D(\tau -1)$.  It is easy to show that the
average size of an avalanche  diverges with exponent $<s> \sim (\Delta
\lambda)^{-\gamma}$ with  $\gamma=1$, and previously we derived that the
critical exponent $\sigma$ describing the cutoff of avalanche sizes
below the critical point is 1/2.  These results verify the scaling
relation $\gamma=(2-\tau)/\sigma$.  The result in Sec.~III for the 
cutoff in the correlation length $\nu=1/2$ in Eq.~(\ref{fscaling}) 
verifies the scaling relation $\nu=1/(\sigma D)$.  Our analysis of
the model in higher than one dimension shows that all of these above
mentioned exponents are independent of dimension.

The punctuated equilibrium behavior, however, does depend on
dimension.  Each site is visited many times in a long lived avalanche.
The intervals between subsequent returns to a given site correspond to
periods of stasis for a given species. As shown in Fig.~\ref{devil},
the accumulated number of returns to a given site forms a ``Devil's
staircase''; the plateaus in the staircase are the periods of stasis
for that species. The punctuations, i. e. the times when the number of
returns increases, occupy a vanishingly small fraction of the time
scale on which the evolutionary activity proceeds.  The distribution
of plateau sizes is the same as the distribution of first returns of
the activity to a given site, $P_{\rm FIRST}(s)$. We found in
dimensions $d\leq 4$ that $P_{\rm FIRST}(s) \sim s^{-\tau_{\rm
FIRST}}$ for large $s$ with $\tau_{\rm FIRST}= 2-d/D$
\cite{scaling}. For $M\to\infty$ in $d=1$ dimension, the scaling
relations therefore predict $\tau_{\rm FIRST}=7/4$. We have measured
$\tau_{\rm FIRST}=1.73\pm 0.05$.

\subsection{Forward Avalanches}

The quantity $P_{\lambda}(s)$ is the conditional probability to have a
forward avalanche of size $s$ given that the signal at the starting
point was equal to $\lambda$.  Such avalanches are defined by looking at
the first moment forward in time when the signal, $\lambda_{min}$
exceeds its current value $\lambda$.  It is important to note that this
conditional forward probability distribution is exactly equal to the
punctuating avalanche distribution.  Punctuating $\lambda$ avalanches
are defined as the intervals separating subsequent punctuations of the
barrier $\lambda$ by the signal $\lambda_{min}(s)$;
see Fig.~\ref{forw}. The probability distribution for the $\lambda$
avalanches does not depend on the precise value of the site that started
the $\lambda$  avalanche (as long as it is $\geq \lambda$)
because this value is erased from the system after the
first update.  Thus,
the minimal numbers selected at different time steps are distributed
as
\begin{eqnarray}
p(\lambda_{min} \geq \lambda)=1/<s>_{\lambda} \nonumber \\
<s>_{\lambda} = \sum_{s=0}^\infty s P_{\lambda}(s) \qquad .
\end{eqnarray}
Substituting Eq. (\ref{z3}) gives $<s>_\lambda= 1/(1-2\lambda)$, and
$p(\lambda_{min}=\lambda)=2$ for $0\leq\lambda\leq 1/2$.

The distribution of {\it all} forward avalanches, $P^{all}_f(s)$ is
obtained by integrating the conditional probability from
Eq.~(\ref{z3}) with the proper weight $p(\lambda_{min}=\lambda)$ for
the starting value of the avalanche \cite{roux,maslov,scaling}.  This
distribution measures avalanches {\it which begin at every time step},
$s'$, and end at the first moment forward in time, $s' +s$, when
$\lambda_{min}(s' +s)> \lambda_{min}(s')$.  We find
\begin{eqnarray}
P^{all}_f(s)& = \int_0^{1/2}p(\lambda_{min}\!=\!\lambda) P_{\lambda}(s)d\lambda
={A(s) \over s} + {1 \over s(2s+1)} \quad ,
\label{eallf}
\end{eqnarray}
which agrees with numerical simulations.
For large $s$ this distribution is a power law
\begin{eqnarray}
P^{all}_f(s) \sim s^{-\tau^{all}_f} 
\quad {\rm where} \quad {\tau^{all}_f}=2 \quad .
\end{eqnarray}
This particular exponent $\tau^{all}_f$ turns out to be
 superuniversal and equals 2 for a wide variety of extremal models
\cite{maslov,scaling}.

\section{Backward Avalanches and Ultrametricity}

The properties of the system under time reversal can be studied in terms
of backward avalanches. Now we look for the first moment {\it back} in
time when \mbox{$\lambda_{min}(s'-s) > \lambda_{min}(s') = \lambda$
\cite{roux,maslov,scaling}.}
The definitions of forward and backward avalanches are illustrated in 
Figs.~\ref{forw} and ~\ref{fback}. These figures demonstrate the hierarchy 
in the avalanche structure: all forward and backward avalanches that start
inside one big forward (backward) avalanche are constrained to not  
go beyond the limits of the parental avalanche and, therefore, can be 
considered to be its sub-avalanches. Each  sub-avalanche in  turn has its 
own sub-avalanches, and so on. One can look at the entire activity as one 
great parental critical avalanche, which began in the distant
past. It contains sub-avalanches of all sizes. 

We can determine the distributions for
$\lambda$ backward avalanches $P_{\lambda}^b(s)$ and all backward
avalanches $P_b^{all}(s)$ exactly.   
Suppose we have a temporal sequence
$\lambda_{min}(s')$ which is an ensemble  of $N$, $\lambda$ punctuating
avalanches, where $N$ is a sufficiently large number. The average number
of $\lambda$ punctuating avalanches of size $s$ in such an ensemble is
given by $N(s)=NP_{\lambda}(s)$. At the end of any $\lambda$ avalanche
of size $s$,  precisely $(s+1)$ new random numbers have been introduced
into the system that were not there when the avalanche began. All these
random numbers are {\it uncorrelated} and uniformly  distributed between
$\lambda$ and $1$ \cite{scaling}. 

To have $\lambda_{min}(s' +s)=\lambda$ we need the minimal number in the
system to lie between $\lambda$ and $\lambda + d\lambda$. 
This number can be only in this set of new random numbers, since at the
beginning of the $\lambda$ avalanche every number in the system was by
definition larger than or equal to $\lambda$. The probability that at
least one of these new  numbers will be between $\lambda$ and $\lambda
+ d\lambda$ is given by $(s+1){d\lambda \over 1-\lambda}$.  Increasing
the parameter to $\lambda + d\lambda$, the number $N$ of avalanches will
be changed by $dN=-N(s+1){d\lambda\over 1-\lambda}$.  Of course the sum
of the temporal durations of these avalanches will remain constant.  
This leads to the following differential equation for the average size
of an avalanche 
\begin{equation}
{d \ln \langle s\rangle_{\lambda} \over d\lambda}= {(s+1) \over 1 -
\lambda}, 
\end{equation}
which is analogous to Eq. (16) in Ref \cite{scaling}, and also shows that
the avalanche size diverges with exponent $\gamma=1$.

The number $N_{\lambda}^b (s)$ of valid $\lambda$ backward  avalanches
of size $s$ in our ensemble of punctuating avalanches is
$N_{\lambda}^b(s)d\lambda =NP_{\lambda}(s) (s+1){d\lambda\over 1-
\lambda}$. Therefore, the conditional probability distribution of
$\lambda$ backward avalanches obeys:
\begin{equation}
\label{e_back}
P_{\lambda}^b(s) = {c(s+1)P_{\lambda}(s) \over (1-\lambda)} = {c  \over
\lambda (1-2\lambda)} {d(\lambda^2 P_{\lambda}(s)) \over d\lambda}
\qquad .
\end{equation}
The proportionality constant $c$ is determined from the normalization
condition $\sum_{s=0}^{\infty} P_{\lambda}^b(s)=1$, so that 
$c= {1 -2\lambda \over 2}$.

The distribution of all backward avalanches measures avalanches which
begin at every time step $s'$, and end at the first moment backward in
time, $s' -s$, when $\lambda_{min}(s' -s) > \lambda_{min}(s')$.  We
find 
\begin{eqnarray}
P^{all}_b(s) &=& \int_0^{1/2}p(\lambda_{min}\!=\!\lambda)
P_{\lambda}^b(s)\,d\lambda
= {\Gamma\left(s+1/2\right) \over 2s\Gamma(1/2) \Gamma\left(s+1\right)} 
+ {1\over 2s(2s+1)}.
\label{allbackward}
\end{eqnarray}
At large times this distribution is a power law $P^{all}_b(s) \sim s^{-
\tau_b^{all}}$ where $\tau_b^{all}=3/2$.  Here we have explicitly
demonstrated that time reversal symmetry is broken. The forward and
backward avalanches have different probability distributions and
different scaling limits at large times.  Eq. (\ref{allbackward}) is in
perfect agreement with numerical simulations.

Unless the backward avalanche has size $s=1$, two extremal values
that are chosen at subsequent times must have both been present when
the first of them was chosen.  These two barrier values will have some
ultrametric distance between them.  This ultrametric distance must be
less than or equal to the backward avalanche size,
because the extremal value that terminates a backward avalanche must
be an ancester to {\it all} of the  extremal values in that backward
avalanche.  The probability to
have an ultrametric distance larger than $s$ is bounded above by
$P^{all}_b(s)$.  We have numerically measured the ultrametric distance
between subsequent activity and find a power law.  In fact, as shown
in Fig. \ref{ultradistances} it has the same leading asymptotic
behavior as $P^{all}_b$ within numerical accuracy.

We conclude by speculating about connections with other phenomena related to
glassy dynamics (see also Ref. \cite{preprint}.
The Directed Polymer in a Random Media (DPRM) \cite{halpin}, which could be
 a paradigm for glassy systems, exhibits an ultrametric structure in
the optimal paths as well as a non-Gaussian tail for the probability
distribution of these paths.  These paths are somewhat analogous to
the activity pattern in our model.   However, unlike the DPRM,
our model is inherently dynamical.   Tang and
Bak \cite{tangbak} found stretched exponential relaxation for the
current in a sandpile model of SOC which could indicate that glassy
dynamics takes place near a critical point.
Recently Stein and Newman
\cite{stein} have
put forward a picture of dynamics on a high dimensional
 rugged fitness landscape based
on an invasion percolation (SOC) picture. 

 Finally, we note our model may
fit into the picture of hierarchically constrained dynamics put
forward by Palmer, Stein, Abrahams, and Anderson \cite{PSAA}.  We have
an equation of motion for the dynamics which takes place in terms of
avalanches spanning all time scales. These avalanches are our
hierarchically constrained degrees of freedom.  Looking at
Fig.~\ref{forw} one notices that every $\lambda$ avalanche is
composed of sub-avalanches which are fully contained within it.  Each
$\lambda$ avalanche cannot terminate until its sub-avalanches finish,
so that the faster degrees of freedom successively constrain slower
ones and form a hierarchy.

\section{Acknowledgements}

This work was supported by the U. S. Department of Energy under Contract
No. DE-AC02-76-CH00016 and DE-FE02-95ER40923.
MP thanks the U.S. Department of Energy Distinguished Postdoctoral
Fellowship Program for financial support.

\appendix{Solution of Eq.~(\ref{E1})}

It is straightforward to solve Eq.~(\ref{E1}) using a generating
function. Defining $p_{\lambda}(x)=\sum_{s=0}^\infty x^s P_{\lambda}(s)$
and applying the initial conditions [see Eq.~(\ref{y1})] $P_\lambda(0)=0$ and
$P_\lambda(1)=(1-\lambda)^2$, we find
\begin{eqnarray}
p_{\lambda}(x)-(1-\lambda)^2=2\lambda(1-\lambda)xp_{\lambda}(x)+
\lambda^2 xp_{\lambda}(x)^2,
\end{eqnarray}
an ordinary quadratic equation for the generating function. Its only
acceptable solution is
\begin{eqnarray}
p_{\lambda}(x)={\left(1-\sqrt{1-4\lambda(1-\lambda)x}\right)^2 \over
4\lambda^2 x}.
\label{px}
\end{eqnarray}
The solution for $P_\lambda(s)$ in Eq.~(\ref{z3}) is simply given by the
coefficients in $x$ of the Taylor series of $p_{\lambda}(x)$. 

The generating function $p_{\lambda}(x)$ has a square-root singularity
which determines the asymptotic behavior of its Taylor coefficients, i.
e. $P_\lambda(s)$. We point out that this asymptotic behavior is a
robust feature with respect of changes in the way the model is updated
at each time step. For instance, if we had not chosen to set the barrier
with the current minimum to unity but to replace it also with a new
random number, we would have obtained a cubic equation for
$p_\lambda(x)$; the solution of which would still be dominated by a
square-root singularity. Furthermore, an update including more than
nearest-neighbor sites would lead to even higher-order algebraic
equations for $p_\lambda(x)$, which are still dominated by the same
square-root singularity. 
It would be interesting to consider changes in the updating rules 
that in fact would replace the leading square-root singularity, and the 
physics that such new rules indicate.

\appendix{Calculation of Eq.~(\ref{Z5})}

Let $r$ be a nonnegative integer. Let $N(i,r)$ be the probability that
a $\lambda_{\rm c}$ avalanche which starts at site $i$ will always be
completely contained inside a box of radius $r$ centered at the origin,
$i=0$. By definition, $N(i,0)\equiv 0$ because in
the initial state the avalanche already contains one barrier which
certainly will never be contained inside a box of vanishing radius. 

The properties of an avalanche starting at $i$ can be related to the
properties of avalanches starting at $i-1$ and $i+1$ by considering all
possible states of the avalanche after the first step. No new active
barriers are created after the first step with probability $(1-
\lambda_{\rm c})^2=1/4$ and the avalanche terminates without ever
spreading beyond the site $i$. A single new active barrier is created
after on time step either to the left or the right of $i$, each with a
probability of $\lambda_{\rm c}(1-\lambda_{\rm c})=1/4$, and $N(i,r)$ is
related to $N(i-1,r)$ or $N(i+1,r)$, respectively. Finally, with a
probability of $\lambda_{\rm c}^2=1/4$ two new active barriers are
created at $i-1$ and at $i+1$, and the avalanche starting at $i$ will
never leave the box if neither avalanche ensuing from $i-1$ and $i+1$
will ever leave the box. Thus, assuming throughout that $r>1$, it is for
$|i|<r-1$
\begin{eqnarray}
N(i,r)={1\over 4}+{1\over 4}\left[N(i-1,r)+N(i+1,r)\right]
+{1\over 4}N(i-1,r) N(i+1,r).
\label{x1}
\end{eqnarray}
Clearly, $N(i,r)\equiv 0$ for all $|i|\geq r$, leading to the boundary
condition for $i=r-1$, 
\begin{eqnarray}
N(r-1,r)={1\over 4}+{1\over 4}N(r-2,r).
\label{x00}
\end{eqnarray}
Since $N(i,r)$ is symmetric in its first argument, we will only consider
nonnegative values of $i$ and obtain another boundary condition at $i=0$
from Eq.~(\ref{x1}): 
\begin{eqnarray}
N(0,r)={1\over 4}+{1\over 2}N(1,r)+{1\over 4}N(1,r)^2.
\label{x01}
\end{eqnarray}

We can simplify Eq.~(\ref{x1}) by substituting
\begin{eqnarray}
N(i,r)=1-f(i,r)
\label{x0.1}
\end{eqnarray}
to obtain for $0<i<r-1$
\begin{eqnarray}
\Delta_i^2 f(i,r)={1\over 2} f(i-1,r) f(i+1,r),
\label{x2}
\end{eqnarray}
where $\Delta_i$ is a difference operator. Eq.~(\ref{x01}) leads to a 
boundary condition at $i=0$,
\begin{eqnarray}
\Delta_i f(i=1,r)={1\over 4}f(1,r)^2,
\label{x2.1}
\end{eqnarray}
while Eq.~(\ref{x00}) gives another boundary condition for $i=r-1$,
\begin{eqnarray}
\Delta_i f(i=r-1,r)=-{3\over 4} f(r-2,r)+{1\over 2}.
\label{x2.2}
\end{eqnarray}
To make further progress we assume $r\gg 1$, which allows to consider
the continuum limit of Eq.~(\ref{x2}). We set $z=i/r$ such that $z$ is
a continuous variable in the unit interval for any $r$ in this limit. We
also set $y(z)=f(i,r)$, where $y$ is a function of $z$ that depends on
$r$ as a parameter. Thus, we can rewrite Eq.~(\ref{x2}) as  
\begin{eqnarray}
{1\over r^2} y''(z)={1\over 2} y(z)^2
\label{x3}
\end{eqnarray}
to leading order in the limit $r\to\infty$. For the boundary conditions
at $z=0$ and $z=1$ we find from Eqs.~(\ref{x2.1}-\ref{x2.2}) 
\begin{eqnarray}
{1\over r} y'(0)={1\over 4} y(0)^2,\qquad {1\over r} y'(1)=-{3\over
4} y(1)+{1\over 2}.
\label{x3.1}
\end{eqnarray}
We note that we can obtain an equation for any value of $\lambda$:
\begin{eqnarray}
{1\over r^2} y''(z)=\left({1\over\lambda}-2\right) y(z)+\lambda y(z)^2 
\end{eqnarray}
which reduces to Eq.~(\ref{x3}) for $\lambda=\lambda_{\rm c}$. It is
easy to show that the linear term dominates on the right-hand side for
$\lambda<\lambda_{\rm c}$, leading to avalanche distributions with an
exponential cut-off for large $r$.

We can integrate Eq.~(\ref{x3}) using standard techniques for autonomous
ordinary differential equations \cite{BeOr}. We set $u(y)=y_r'(z)$, use
the chain rule to get $y_r''(z)=u'(y)u(y)$, and integrate once to find 
\begin{eqnarray}
{1\over r} y'(z)=\pm \sqrt{{1\over 3}y(z)^3+C}.
\end{eqnarray}
Since $y(z)$ is a rising function of $z$, we choose the positive root.
The integration constant $C$ can be rewritten using the boundary
condition at $z=0$ in Eq.~(\ref{x3.1}) as $C=-{1\over 3} y(0)^3$, where
we neglected terms  of higher order in $y(0)$ because $y(0)$ is expected
to be small for  $r\to\infty$.

Integrating one more time we obtain
\begin{eqnarray}
\int_1^{y(z)\over y(0)} {d\zeta\over\sqrt{\zeta^3-1}}= z\,
r\sqrt{y(0)\over 3}. 
\end{eqnarray}
We find at $z=1$, using Eq.~(\ref{x3.1}), that $y(1)=2/3$, because
$y'(1)/r$ can be shown to be negligible. Thus, $y(1)/y(0)\to\infty$, and
we  obtain Eq.~(\ref{Z5})
\begin{eqnarray}
f(0,r)\sim y(0)\sim {3\over r^2}\left(\int_1^\infty
{d\zeta\over\sqrt{\zeta^3-1}}\right)^2={1\over 3}\left[{\Gamma({1\over
6})\Gamma({1\over 2})\over\Gamma({2\over 3})}\right]^2{1\over
r^2}\approx {17.69\over r^2}.
\label{x4}
\end{eqnarray}
Note that this result verifies our assumption that $y(0)$ is small.
Using dominate balance techniques \cite{BeOr} we can show that this
solution is in fact the only consistent solution.

\appendix{Leading Asymptotic Behavior of the Spatio-Temporal
Correlations with Respect to a Particular Site}

The nonlinear integro-difference equation.~(\ref{envelope}) can be solved
exactly in the continuum limit. Taking the continuum limit is justified
because we are ultimately interested in the behavior of $F(r,s)$ for
sufficiently large values of $r$ and $s$. In general, to obtain the
asymptotic behavior of a difference equation from the corresponding
differential equation can be tricky even in the linear case \cite{BeOr}.
But comparison with our calculation in Appendix D, where the continuum
limit poses no problem, independently confirms our approach here.

With $f(r,x)=\sum_{s=0}^\infty x^s F_\lambda(r,s)$ and 
$p(x)=\sum_{s=0}^\infty x^s P_\lambda(s)$ as generating functions, we obtain
from Eq.~(\ref{envelope}), using Eq.~(\ref{px}),
\begin{eqnarray}
&\Delta_r^2 f(r,x)=A(x) f(r,x) +B(x) f(r-1,x) f(r+1,x),&\cr
\noalign{\medskip}
&f(0,x)=p(x),\quad {\rm and}\quad f(\infty,x)=0,&
\label{y3}
\end{eqnarray}
defining
\begin{eqnarray}
A(x)={1-2x\lambda(1-\lambda)-2x\lambda^2p(x)\over x\lambda(1-
\lambda)+x\lambda^2 p(x)},\quad B(x)={x\lambda^2\over x\lambda(1-
\lambda)+x\lambda^2 p(x)}.
\label{y3.1}
\end{eqnarray}

We can take the continuum limit of Eq.~(\ref{y3}) and get to leading
order for large $r$ an ordinary second-order nonlinear differential
equation for $f$ as a function of $r$:
\begin{eqnarray}
f(r)''=A f(r)+B f(r)^2,\quad f(0)=p,\quad f(\infty)=0,
\label{y4}
\end{eqnarray}
where we have suppressed dependence on the parameter $x$. Using again
the techniques for autonomous equations (see Appendix B) and the fact
that $f(\infty)=0$, Eq.~(\ref{y4}) can be solved exactly to give
\begin{eqnarray}
f(r)=p \left[\cosh\left({1\over 2}\sqrt{A} r\right)+\sqrt{1+{2Bp\over
3A}}\sinh\left({1\over 2}\sqrt{A}r\right)\right]^{-2}. 
\label{y4.1}
\end{eqnarray}
Thus, we obtain from Eq.~(\ref{y4.1}) a closed-form expression for the
envelope function of the spatio-temporal distribution of avalanches: 
\begin{eqnarray}
F_\lambda(r,s)\sim -\oint {dx\over 2\pi i} x^{-s-1} f(r,x),
\label{y5}
\end{eqnarray}
where the contour encircles a small neighborhood of the origin in the
complex $x$-plane in the positive direction. 

{}From now on, we only consider the critical case, $\lambda=\lambda_{\rm
c}=1/2$. The integrand in Eq.~(\ref{y5}) can be expanded for large $s$
in the neighborhood of $x=1$ by substituting $x=1-u/s$. Then, the
integration for $u$ follows a contour that crosses the positive real
axis from above near the origin in the complex-$u$ plane.  With
\begin{eqnarray}
p(x)\sim 1-2\sqrt{u\over s}+2{u\over s},\quad A\sim 2\sqrt{u\over
s}+2{u\over s},\quad B= {1\over 2}+{1\over 2}\sqrt{u\over s},
\end{eqnarray}
we find
\begin{eqnarray}
F(r,s)\sim \int_{\cal C} {du\over 2\pi is} e^u f\left(r,1-{u\over s}\right),
\end{eqnarray}
where an analysis of Eq.~(\ref{y4.1}) yields
\begin{eqnarray}
f(r,1-{u\over s})\sim\cases{{12\over r^2}
\left[-\left({u\over s}\right)^{1\over 2}\left({r^2\over 6}+1
\right)+\left({u\over s}\right)^{3\over 2}{r^6\over 756}+\ldots\right]
\quad&$\left(1\ll r^4\ll s\right)$,\cr
\noalign{\medskip}
24 \left({u\over s}\right)^{1\over 2}\exp\left[-\sqrt{2}r \left({u\over
s}\right)^{1\over 4}\right]\quad&$\left(s\ll r^4\right)$.} 
\label{y6.1}
\end{eqnarray}
In the first case of Eq.~(\ref{y6.1}) we have neglected terms with
integer powers in $u$ which would vanishes in the following integration.

{}For $r^4\gg s$, we evaluate the integral for $F(r,s)$ by steepest-descent
analysis similar to Sec.~III, but taking account also of the non-
exponential factor in the integrand.  For $1\ll r^4\ll s$, we use
Hankel's contour integral representation of the $\Gamma$-function
\cite{A+S}
\begin{eqnarray}
{1\over\Gamma(-\nu)}=-\int_{\cal C} {du\over 2\pi i} e^u u^\nu
\end{eqnarray}
to find
\begin{eqnarray}
{}F(r,s)\sim\cases{{1\over \sqrt{\pi}} s^{-{3\over 2}}\left(1+{6\over
r^2}+{1\over 84}{r^4\over s}+\ldots\right) \quad&$(1\ll r^4\ll s)$,\cr
\noalign{\medskip}
\sqrt{24\over\pi} s^{-{3\over 2}}\left(r^4\over s\right)^{1\over 3}
\exp\left[-{3\over 4}\left({r^4\over s}\right)^{1\over 3}\right]
\quad&$(r^4\gg s\gg 1)$,\cr} 
\label{h10}
\end{eqnarray}
where the second case is our main result for the non-Gaussian tail given
in Eq.~(\ref{nongaussian}).

\appendix{Leading Asymptotic Behavior of the Spatio-Temporal
Correlations with Respect to a Box}

In this Appendix we calculate the distribution for the size of a box in
space and time such that a $\lambda$-avalanche will be fully contained
in it. This notion, which easily generalizes to higher dimensions,
extends on the calculation in Sec.~III and Appendix C, where we only
considered avalanches with respect to a single site. While the
calculation here (which is similar to Appendix B) is somewhat more
extensive, the results are virtually identical and justify our
simplified approach in Sec.~III.

Let $P(r,s,i)$ be the probability for a $\lambda$-avalanche, which
starts at time $s=0$ with a single active barrier on site $i$, to have
no active barriers for the first time at time $s$, and to not have left
a box $i\in (-r,r)$. We merely consider avalanches in the critical state
and set $\lambda=\lambda_{\rm c}=1/2$. By definition, $P(r=0,s,i)\equiv
0$ for all $s\geq 0$. Furthermore, $P(r,s=0,i)\equiv 0$ for all $r$,
because by definition no avalanche ends at time $s=0$. Ultimately, we
want to find the distribution $F(r,s,0)$ of avalanches of duration $s$
that start at the origin $i=0$ and are completely contained in a box of
radius $r$. A generic avalanche is plotted in Fig.~\ref{box}. The smallest 
box it is fully contained in is of size $r=18$ in this case.
In correspondence with Eq.~(\ref{deffrs}) it is
\begin{eqnarray}
F(r,s,i)=P(r=\infty,s,i)-P(r,s,i),
\label{reflect}
\end{eqnarray}
where $P(r=\infty,s,0)=p(x)$ is given in Eq.~(\ref{px}).

As before, the properties of an avalanche that originates at $s=0$ can
be deduced from the properties of avalanches that ensue after the first
update. The original avalanche can either terminate after the first
update when the update does not produce any new active barriers with
probability $(1-\lambda)^2=1/4$, or it can generate new avalanches by
creating new active barriers. If the first update creates exactly one
new barrier with likelihood $\lambda(1-\lambda)=1/4$ either to the left
or to the right of site $i$, the properties of an original avalanche of
duration $s$ is related to the properties of an avalanche of duration
$s-1$ with regard to a site of distance $i-1$ or $i+1$, respectively. If
the first update creates two new active barriers with probability
$\lambda^2=1/4$ to the left and the right of site $i$, two new
avalanches ensue. Then, the properties of the original avalanche of
duration $s$ is related to the properties of all combinations of two
avalanches of combined duration $s-1$. For any such combination, the
probability to not leave the box when starting at site $i$ is given
simply by the product of the probabilities for the two ensuing
avalanches to not leave the box after starting at site $i-1$ or $i+1$,
respectively. We thus obtain for $r\geq 1$ and $|i|<r$ that
$P(r,s=1,i)=1/4$, and for all $s\geq 1$ that
\begin{eqnarray}
P(r,s+1,i)&=&{1\over 4}\left[P(r,s,i-1)+P(r,s,i+1)\right]\cr
\noalign{\medskip}
&&\quad +{1\over 4} \sum_{s'=0}^{s} P(r,s',i-1) P(r,s-s',i+1).
\label{y7}
\end{eqnarray}
Since $P(r,s,i)$ is symmetric in $i$, we restrict ourselves to
nonnegative values of $i$, leading to a boundary condition at $i=0$:
\begin{eqnarray}
P(r,s+1,0)={1\over 2} P(r,s,1)+{1\over 4} \sum_{s'=0}^{s} P(r,s',1)
P(r,s-s',1)\quad (s\geq 1,r\geq 1).
\end{eqnarray}
Since $P(r,s,i\geq r)\equiv 0$, we obtain a second boundary condition at
$i=r-1$:
\begin{eqnarray}
P(r,s-1,r-1)={1\over 4}P(r,s,r-2)\quad (s\geq 1,r\geq 2). 
\label{y8} 
\end{eqnarray}

Using Eq.~(\ref{reflect}), and defining 
$f(r,x,i)=\sum_{s=0}^\infty x^s F(r,s,i)$, we obtain for $1\leq i\leq r-2$
\begin{eqnarray}
\Delta_i^2 f(r,x,i)=A(x) f(r,x,i) +B(x) f(r,x,i-1) f(r,x,i+1)
\label{y10}
\end{eqnarray}
where $A(x)$ and $B(x)$ are the same as in Eq.~(\ref{y3.1}),
supplemented by the boundary conditions
\begin{eqnarray}
\Delta_i f(r,x,i=1)&=&\left[1-{x\over 4}-{x\over 4}p(x)\right] f(r,x,1)
+{x\over 4} f(r,x,1)^2,\cr
\noalign{\medskip}
\Delta_i f(r,x,i=r-1)&=&\left({x\over 4}-1\right) f(r,x,r-2)
-p(x)\left(1-{x\over 4}\right)-{x\over 4}.
\end{eqnarray}

As before in Appendix C, we expand the equations in the limit $x\to 1$
to analyze the avalanche distribution for large times $s$. Then
Eq.~(\ref{y10}) simplifies for $1\leq i\leq r-2$ to
\begin{eqnarray}
\Delta_i^2 f(r,x,i)\sim 2\sqrt{1-x} f(r,x,i)+{1\over 2} f(r,x,i-1) f(r,x,i+1)
\label{y11}
\end{eqnarray}
with the boundary conditions
\begin{eqnarray}
\Delta_i f(r,x,i=1)&\sim&\sqrt{1-x}f(r,x,1) +{1\over 4} f(r,x,1)^2,\cr
\noalign{\medskip}
\Delta_i f(r,x,i=r-1)&\sim&-{3\over 4} f(r,x,r-2) +{1\over 2}.
\end{eqnarray}
{}For sufficiently large $r$, we can take the continuum limit of these
equations where $f(r,x,i)\to y(z)/r^2$ with $z=i/r$ as a continuous
variable in the unit interval. Eq.~(\ref{y11}) then approaches
\begin{eqnarray}
y''(z)= 2r^2\sqrt{1-x} y(z)+{1\over 2} y(z)^2,\quad
y'(0)=r\sqrt{1-x} y(0),\quad y(1)={2\over 3}r^2.
\label{y12}
\end{eqnarray}
We can obtain a first integral of Eq.~(\ref{y12}) using again the
technique for autonomous equations:
\begin{eqnarray}
y'(z)=\pm\sqrt{{1\over 3}\left[y(z)^3-y(0)^3\right]+2r^2(1-x)^{1\over 2}
\left[y(z)^2-y(0)^2\right]},
\end{eqnarray}
assuming that $y'(0)\to 0$ for $r\to\infty$ sufficiently fast. Since
$y(z)$ is a rising function of $z$, we have to chose the positive root.
Integrating again, and using $y(1)$, we obtain 
\begin{eqnarray}
\int_1^{2r^2\over 3y(0)}{d\zeta\over \sqrt{\zeta^3-1
+\alpha\left(\zeta^2-1\right)}}\sim\sqrt{y(0)\over 3}~z
\quad {\rm with} \quad\alpha={6r^2\sqrt{1-x}\over y(0)}.
\label{integral}
\end{eqnarray}
{}For $y(0)\ll r^2\sqrt{1-x}$, i. e. $\alpha\gg 1$, we get
\begin{eqnarray}
\sqrt{y(0)\over 3}\sim {1\over\sqrt{\alpha}} \ln\left[
{3\over2}\alpha \right],
\end{eqnarray}
which yields
\begin{eqnarray}
f(r,x,0)\sim {y(0)\over r^2}\sim 3\sqrt{1-x}\exp\left(-
\sqrt{2}  r(1-x)^{1\over 4}\right)\quad (1\ll (1-x)\ll r^4), 
\label{y14}
\end{eqnarray}
and by steepest-descent analysis as before
\begin{eqnarray}
F(r,s,0)\sim C(r,s) \exp\left[-{3\over 4} 
\left({r^4\over s}\right)^{1\over 3}\right],
\end{eqnarray}
confirming the non-Gaussian tail found in Eq.~(\ref{nongaussian}).
A similar consideration of the
integral in Eq.~(\ref{integral}) would determine the behavior for
$r^2\gg y(0)\gg r^2\sqrt{1-x}$, i. e. $\alpha\ll 1$.

\figure{\label{devil}
Punctuated equilibrium behavior for the evolution of a single  species
in the one-dimensional $M=\infty$ model.  The vertical axis is the
total number of returns of the activity to some site as a function of
time $s$.  Note the presence of plateaus (periods of stasis)  of all sizes.
The distribution of plateau sizes scales as $s^{-7/4}$.}
\epsfxsize=300pt
\epsfysize=400pt
\epsffile{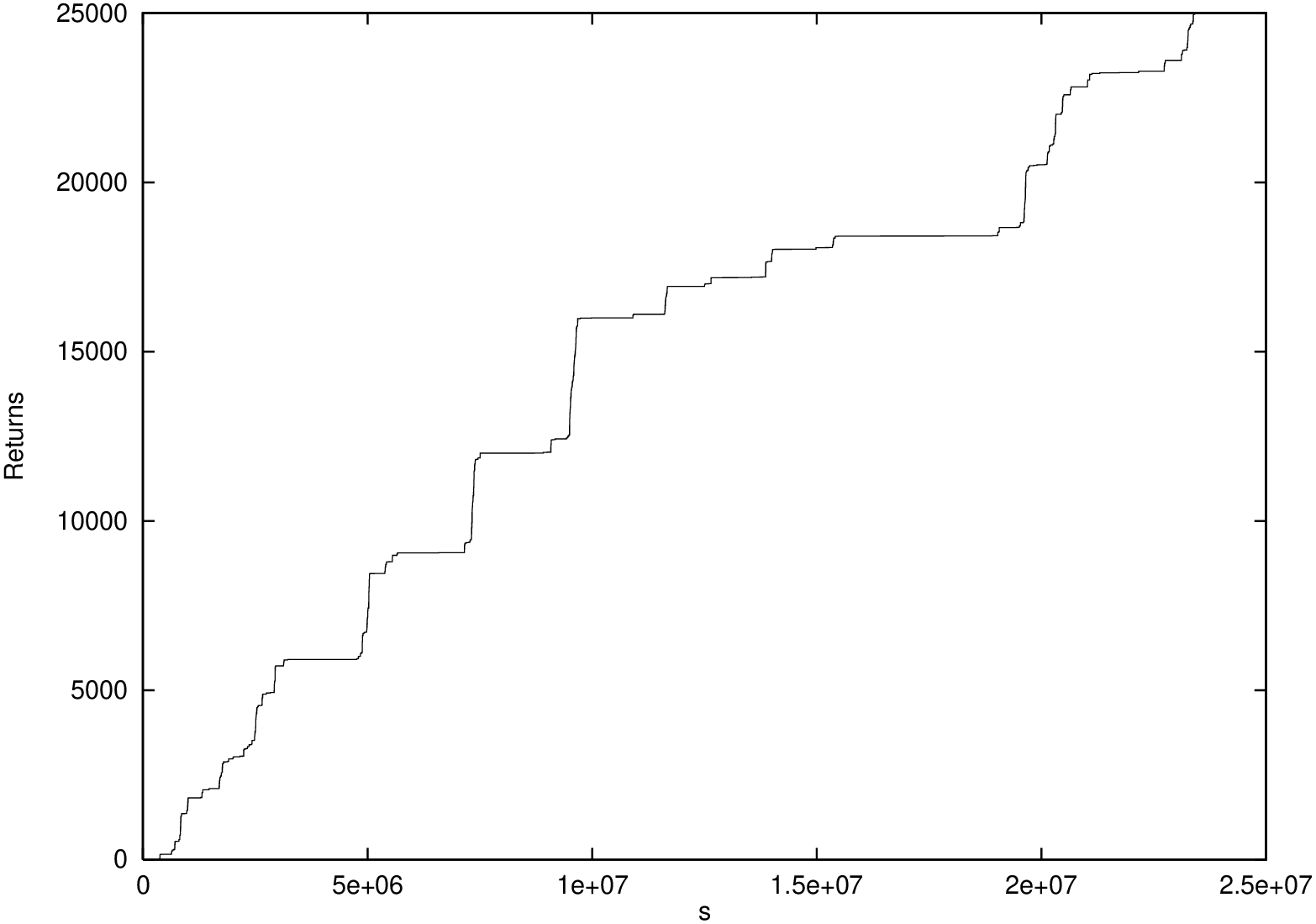}
\newpage

\figure{\label{stationary}
Snapshot of the stationary state in a finite one dimensional system for
the $M=1$ (Bak-Sneppen) model.
Except for the avalanche which consists of small fitness values in 
a localized region, almost all the fitness values in the system are 
above a self-organized threshold $\lambda_c$.}
\epsfxsize=350pt
\epsfysize=450pt
\epsffile{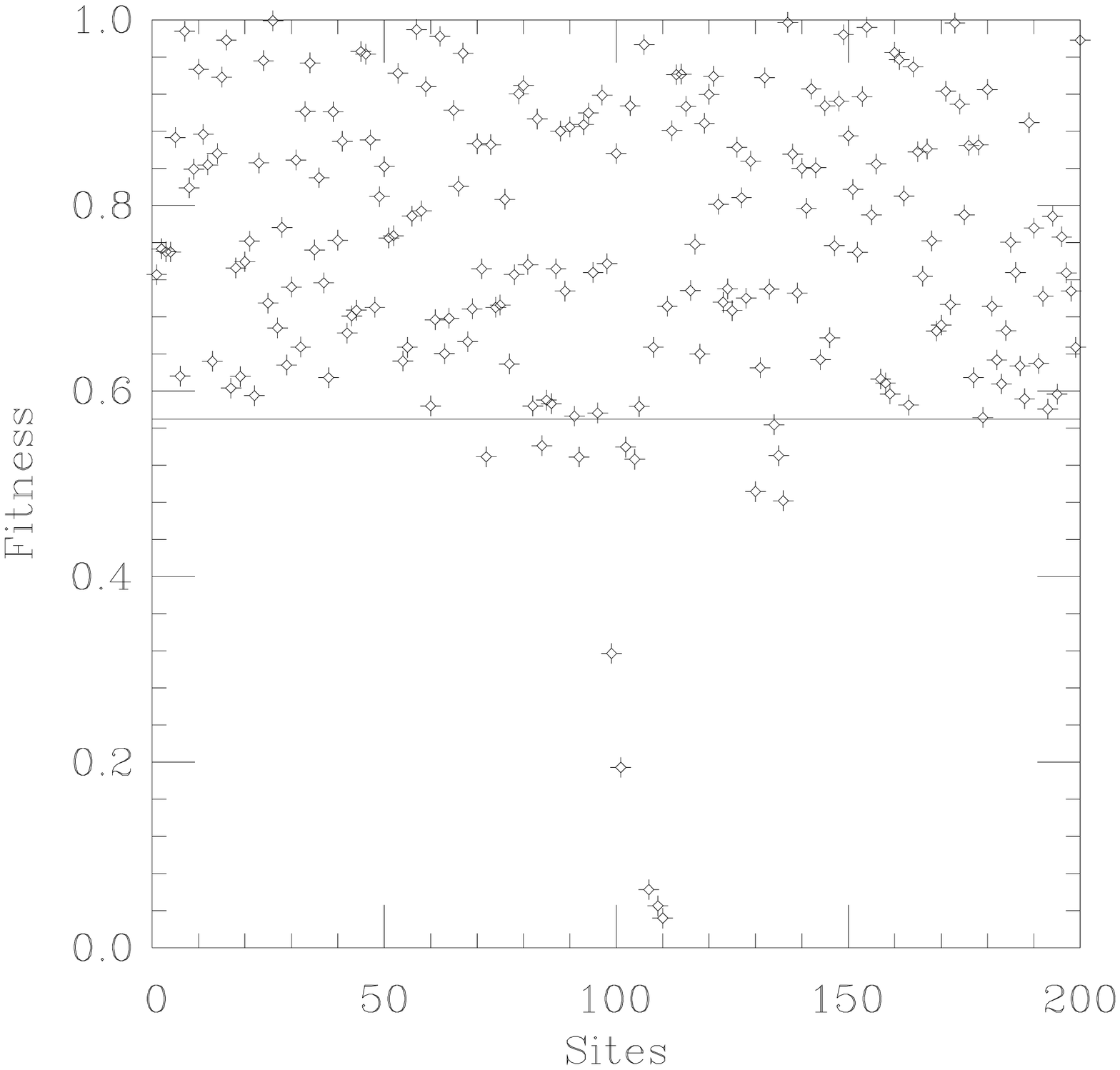}
\newpage

\figure{\label{ultratree}
Ultrametric tree structure.  At any given time, indicated by the vertical
axis, all of the active sites below threshold 
have an ancestry which forms a tree.
The ultrametric distance between any pair is the distance back in time
to the first common ancestor.}
\epsffile[0 600 100 800]{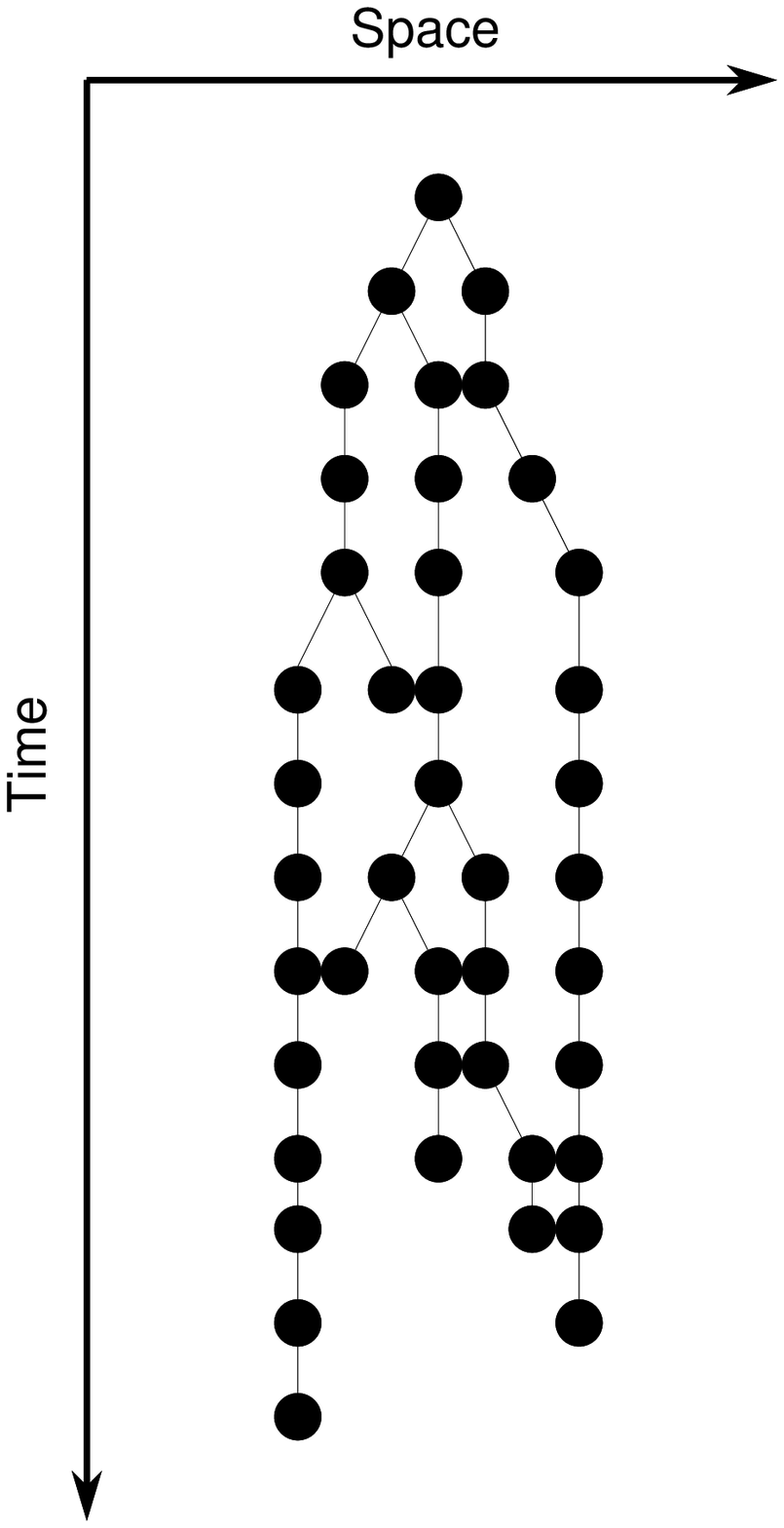}
\newpage

\figure{\label{f.box}
Numerical verification of Eq.~(\ref{Z5}). We sampled the
probability for an avalanche to start at the origin and to be
contained by the smallest possible box of radius $r$ for $r\leq 250$
with increments of $\Delta r=10$. Squares represent the measured
values from a sample of $2\times10^6$ avalanches for this probability,
corresponding to $-\partial_r f(0,r)\sim 35.48\ldots r^{-3}$ which is
drawn as a solid line.  }
\epsfxsize=350pt
\epsfysize=450pt
\epsffile{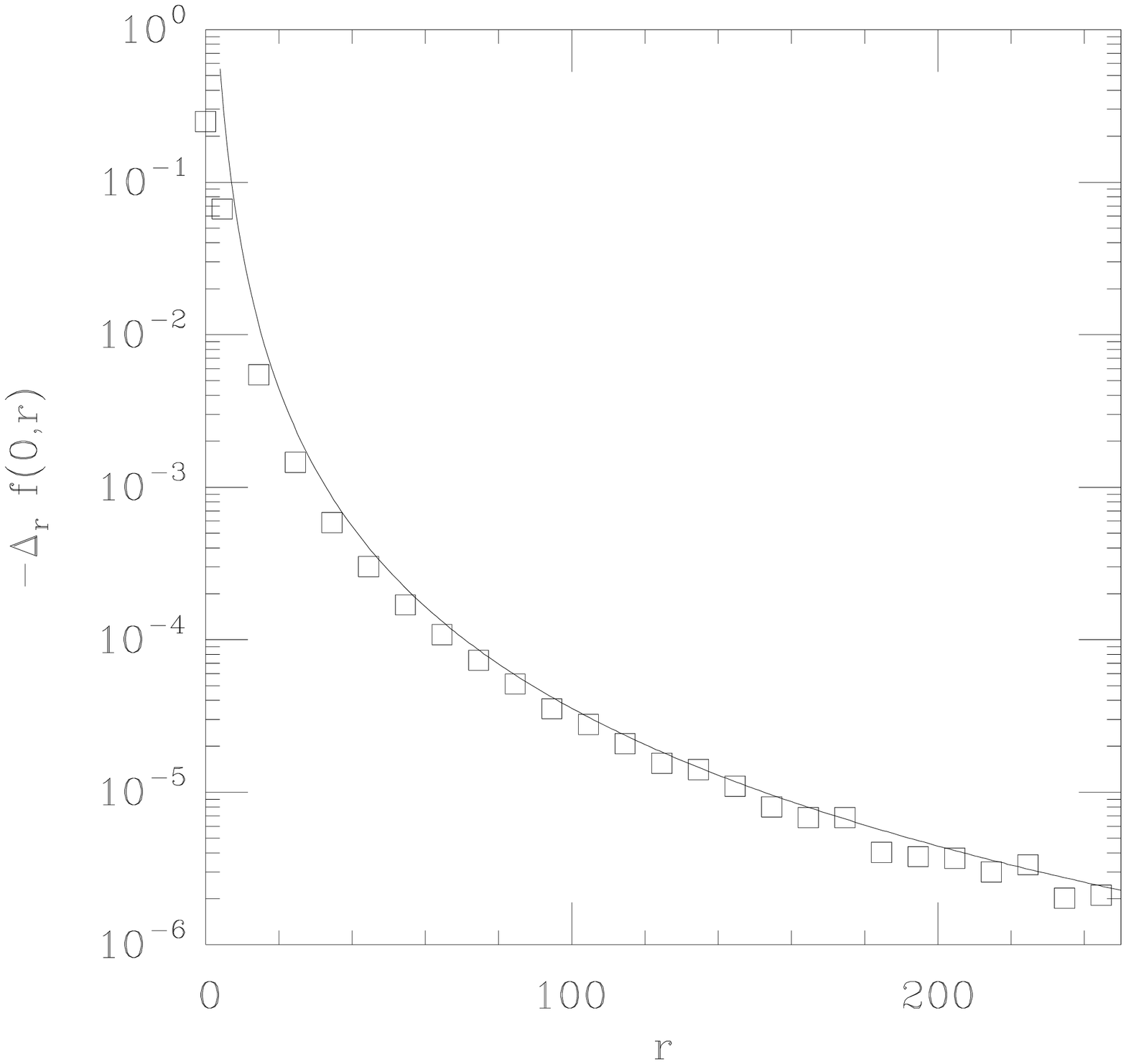}
\newpage

\figure{\label{box}
Plot of a typical $\lambda_{\rm c}$ avalanche for $M=\infty$, starting 
at the origin at time $s=0$, and ending at time $s=976$. At each time 
step, every site with at least one active barrier is marked 
$\diamondsuit$, and the sequence 
of minimal sites is connected with a line showing jumps of various sizes.
Apparent also are the punctuated equilibria for each site which extend
over many sizes. The propagator $G(r,s)$ of the activity is the 
probability to have a minimum at site $r$ at time $s$, while $F(r,s)$ is
the probability for an avalanche to end at $s$ (here: $=976$) and
to have reached a particular site $r$ during its duration.}
\epsfxsize=300pt
\epsfysize=400pt
\epsffile{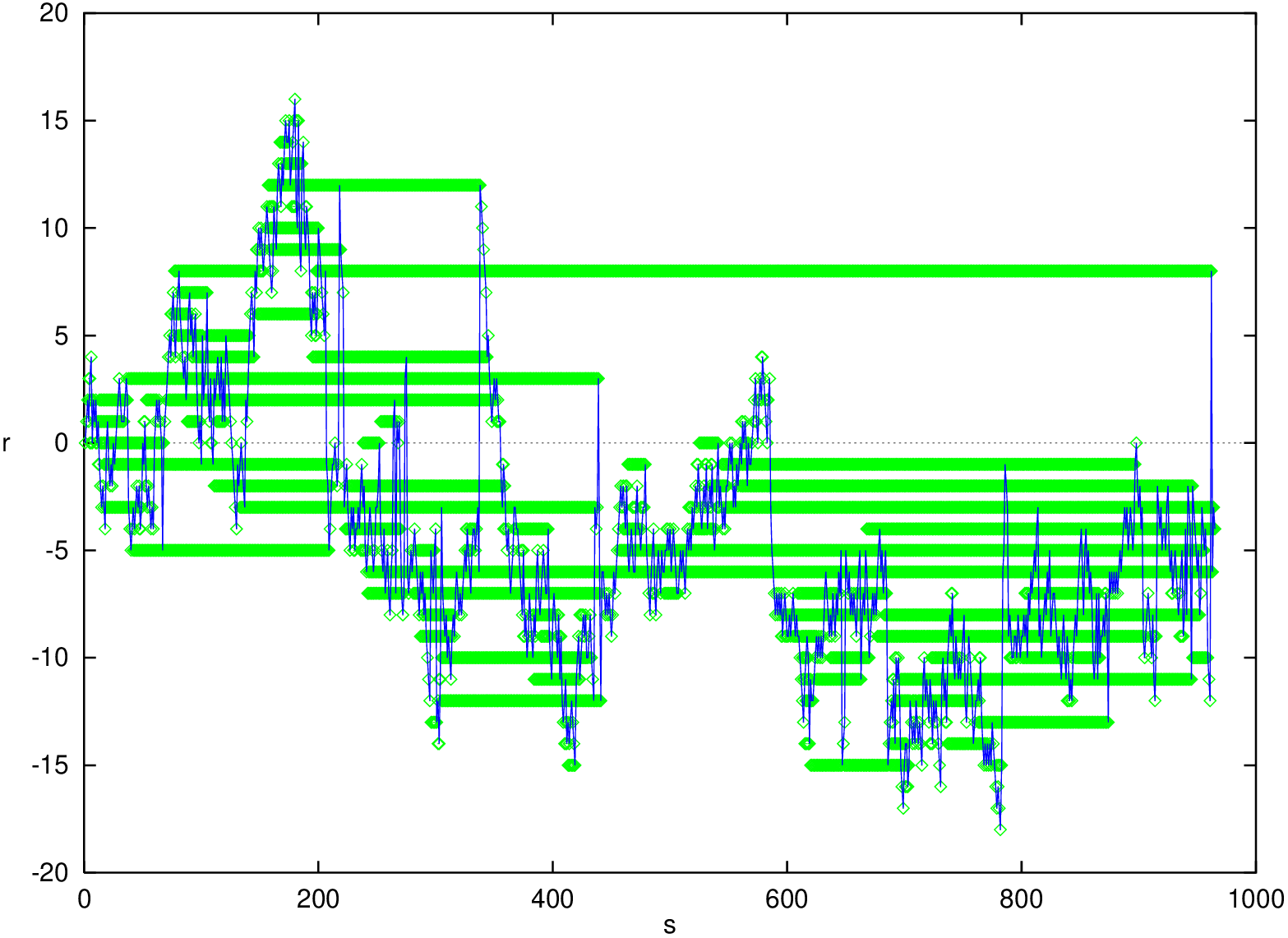}
\newpage

\figure{\label{scaling}
Counting rates for the activity plotted as a function of time $s$. Each
curve is labeled in descending order by $i=\lceil\log_2 x\rceil$ where 
$x=r^4/s$ is the scaling variable [see Eq.~(\ref{gx})]. Data with $i>12$ 
has been disregarded due to insufficient statistical accuracy.}
\epsfxsize=350pt
\epsfysize=450pt
\epsffile{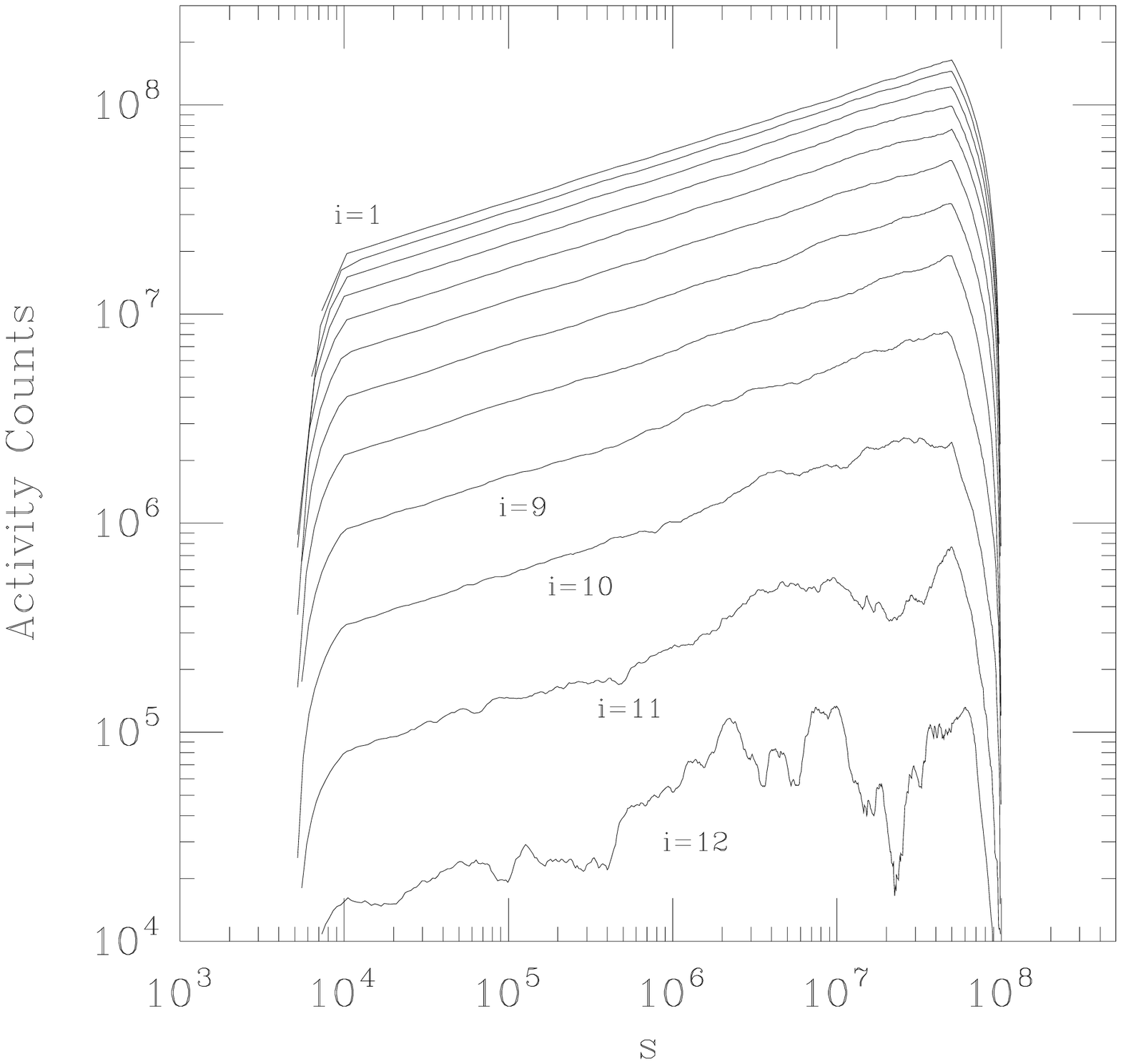}
\newpage

\figure{\label{collapse}
The normalized distribution of activity $G(r,s)$ derived from 
Fig.~\ref{scaling}.  Each value of $G(r,s)$ is divided by its
corresponding value of $r$ and cut off below $s=10^4$ and above $s=5\times
10^7$. All data ``collapses'' onto plateaus with a mean value 
for each plateau given by
$\times$  on the right, again labeled by 
$i=\lceil\log_2 x\rceil$. The standard deviation for
the data in each plateau is indicated by error bars. }
\epsfxsize=350pt
\epsfysize=450pt
\epsffile{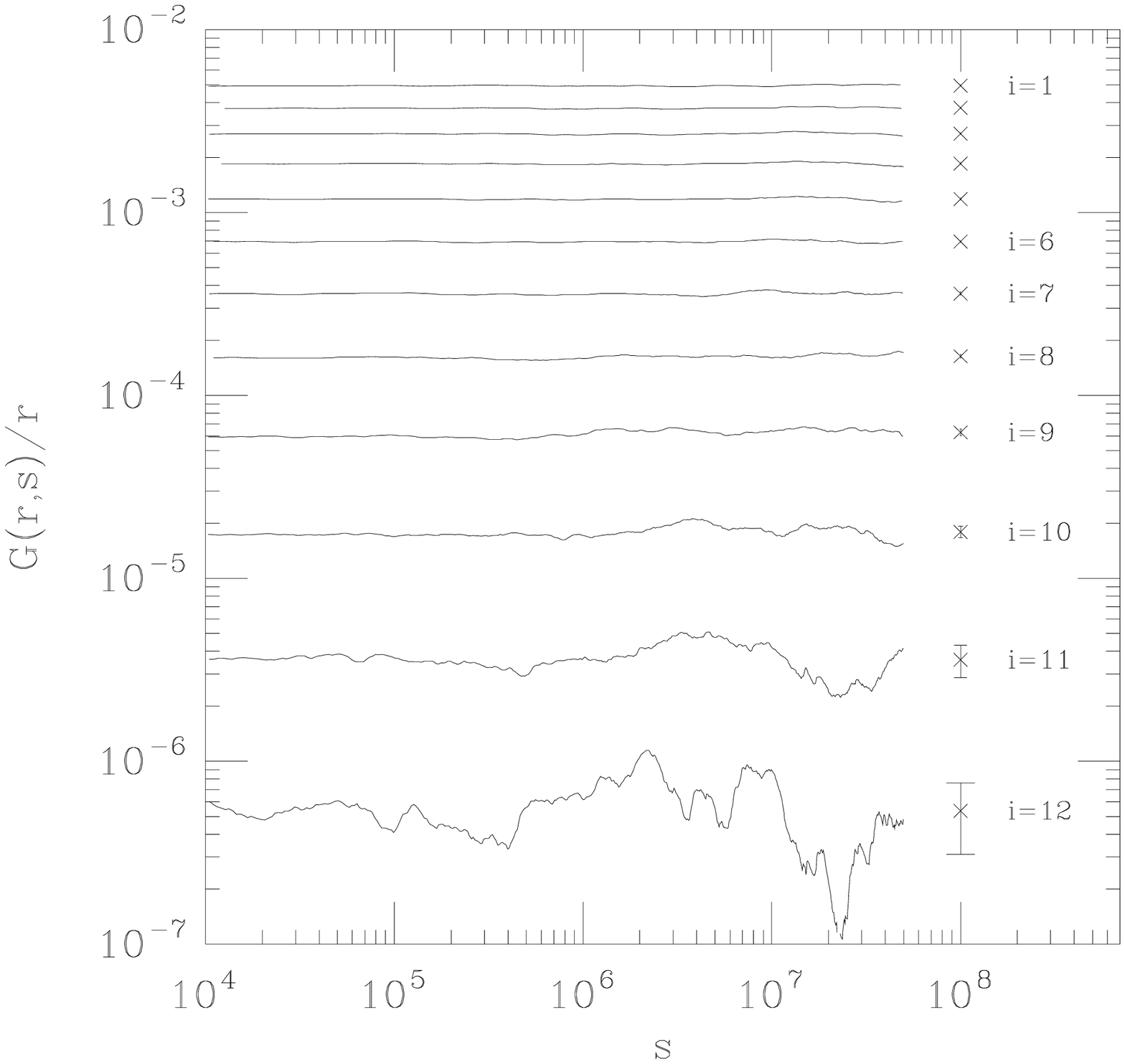}
\newpage

\figure{\label{extrapolation}
Plot of the extrapolants from Eq.~(\ref{extraseq}) for each $i=\lceil\log_2
x\rceil$
as function of $1/\ln x$. Asymptotically for $1/\ln x\to 0$, this
sequence approaches the value of $\alpha$. From the extrapolation of the
sequence to $x\to\infty$ (indicated by the lines that extend the
sequence and its errors to the ordinate) we
estimate $\alpha=0.35\pm 0.03$, in good agreement with $\alpha=1/3$.}
\epsfxsize=350pt
\epsfysize=450pt
\epsffile{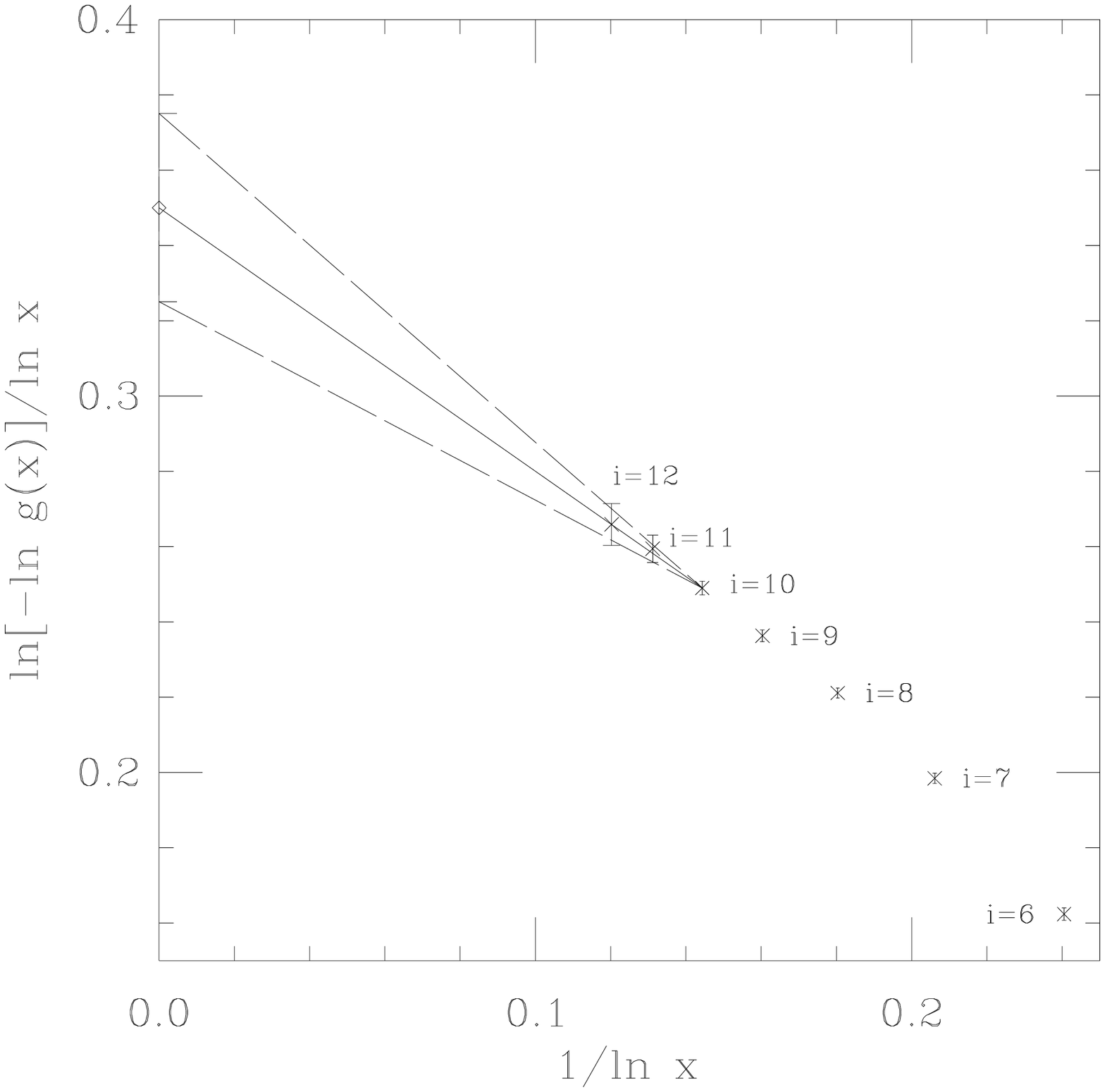}
\newpage

\figure{\label{compact}
Plot of the mean  width $R(s)$ of the compact region 
of covered sites for surviving avalanches at time $s$ for $s<10^5$.
The result of a simulation involving $10^8$ avalanches is given by
$\diamondsuit$, and the solid line is given by 
$6.10\left(s^{1\over 4}-1.0\right)$, obtained  
from an asymptotic fit of the data for large $s$.}
\epsfxsize=300pt
\epsfysize=400pt
\epsffile{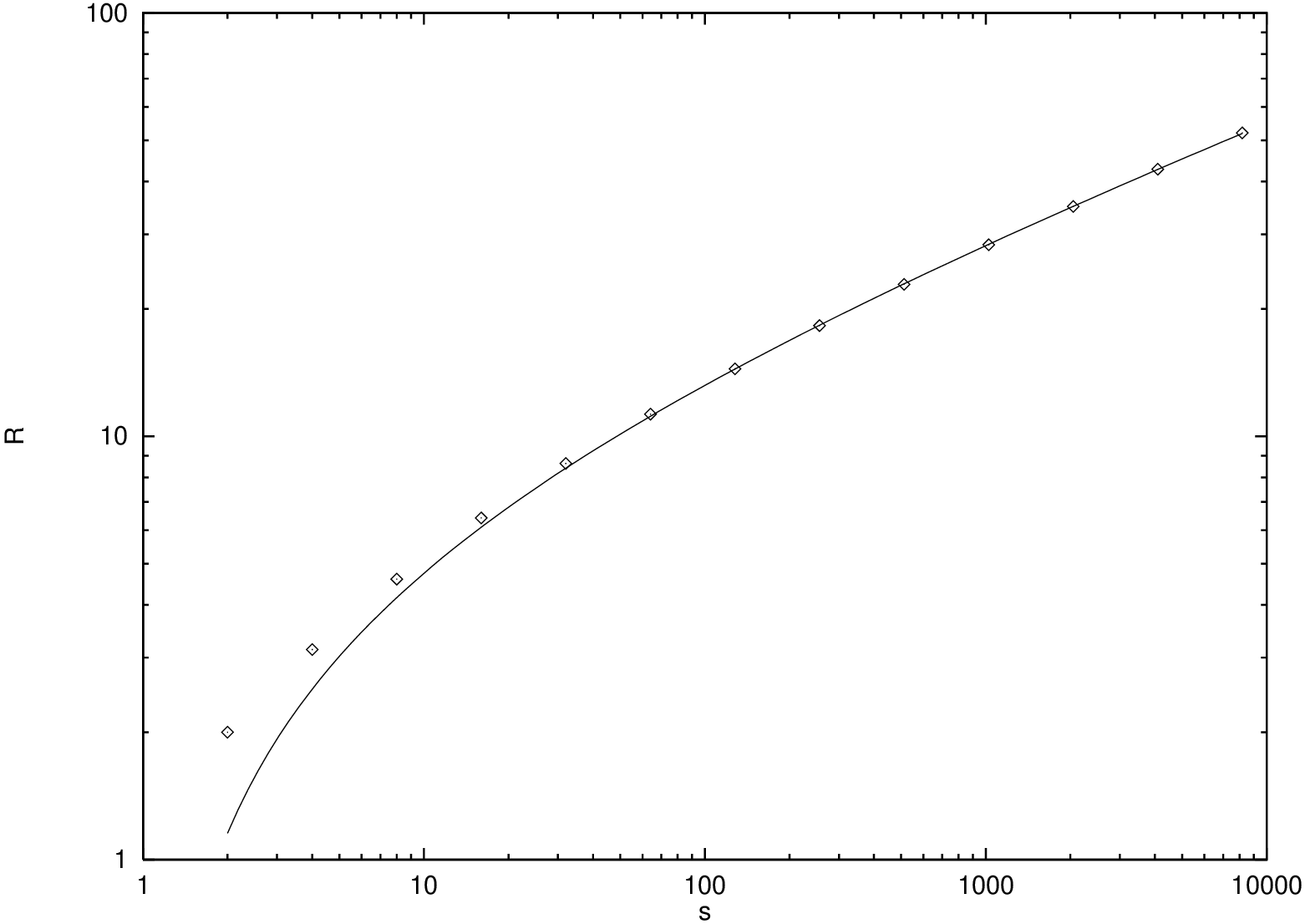}
\newpage

\figure{\label{forw}
Plot of a part of the sequence of minimal random numbers 
$\lambda_{\rm min}(s)$ chosen for an update at time $s$ in a 
$\lambda_{\rm c}$ avalanche for $M=\infty$. The durations of a hierarchy of 
$\lambda$ avalanches is indicated by forward arrows, where 
$\lambda=\lambda_{\rm min}(s)$. Longer avalanches with larger values of
$\lambda$ contain many shorter
avalanches which have to finish before the longer avalanche can terminate.
Note that an update punctuates any $\lambda$ avalanche with
$\lambda\leq \lambda_{\rm min}(s)$.}
\epsfxsize=350pt
\epsfysize=450pt
\epsffile{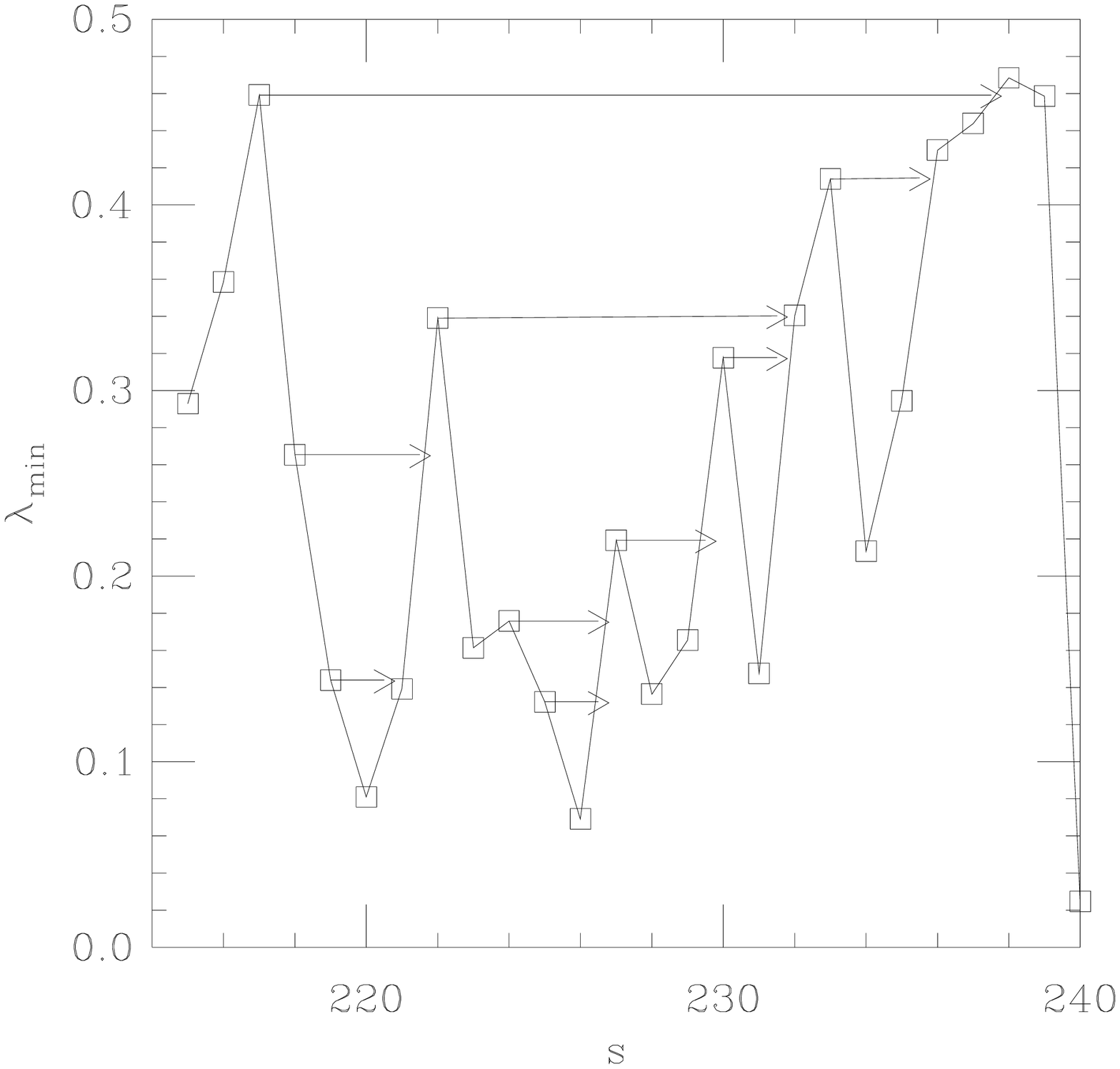}
\newpage

\figure{\label{fback}
Same sequence $\lambda_{\rm min}(s)$ as in Fig.~\ref{forw}, where the
durations of backward $\lambda$ avalanches is indicated by backward
arrows. A similar hierarchical structure of sub-avalanches emerges,
although with a different distribution than for forward $\lambda$
avalanches in reflection of the irreversibility of the update process.}
\epsfxsize=350pt
\epsfysize=450pt
\epsffile{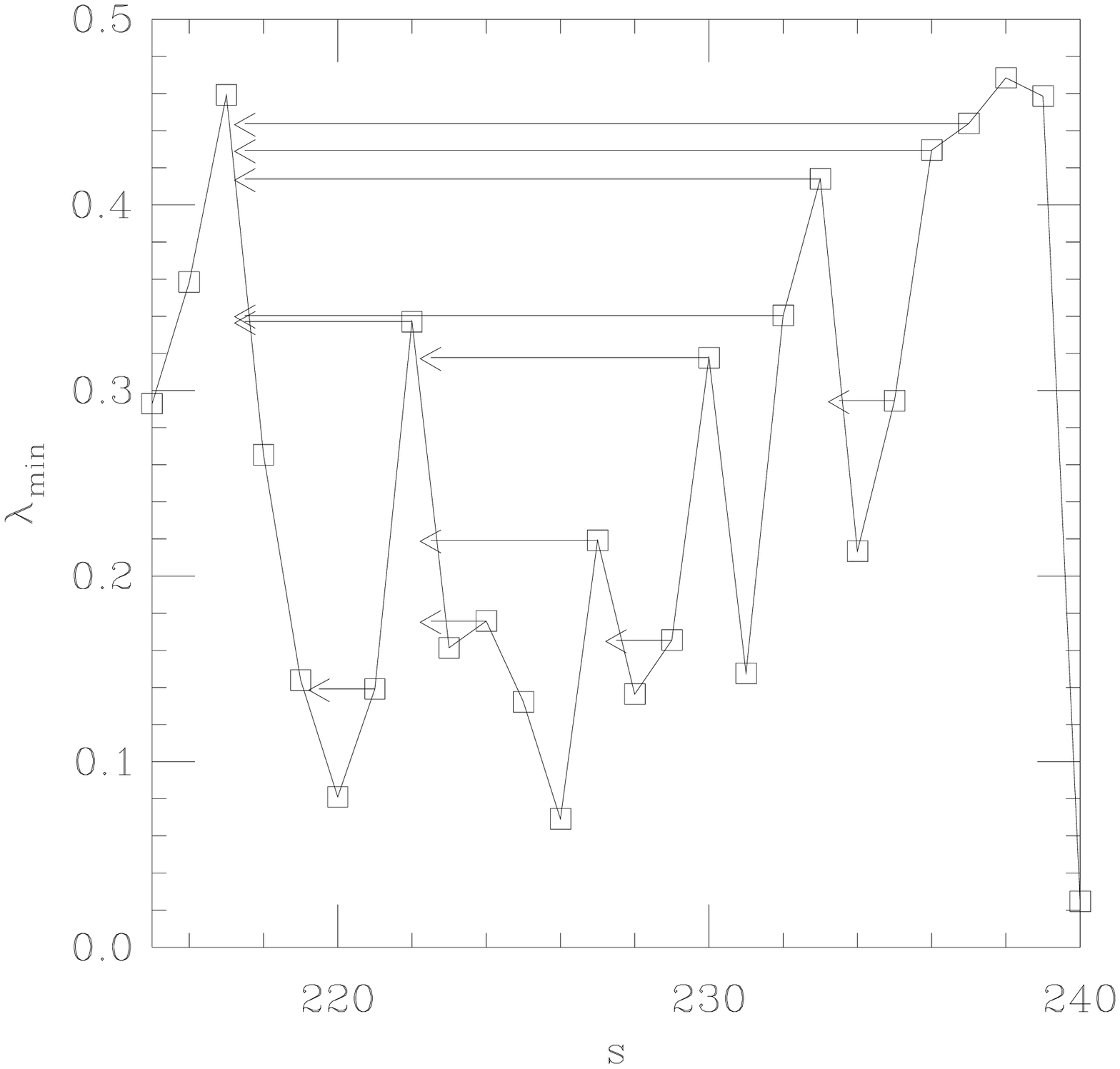}
\newpage

\figure{\label{ultradistances}
Log-log plot of the distribution for the duration of backward avalanches
($+$) and the distribution for the ultrametric distances between
subsequent minimal sites ($\times$). Both functions seem to coincide
asymptotically for large $s$.}
\epsfxsize=350pt
\epsfysize=450pt
\epsffile{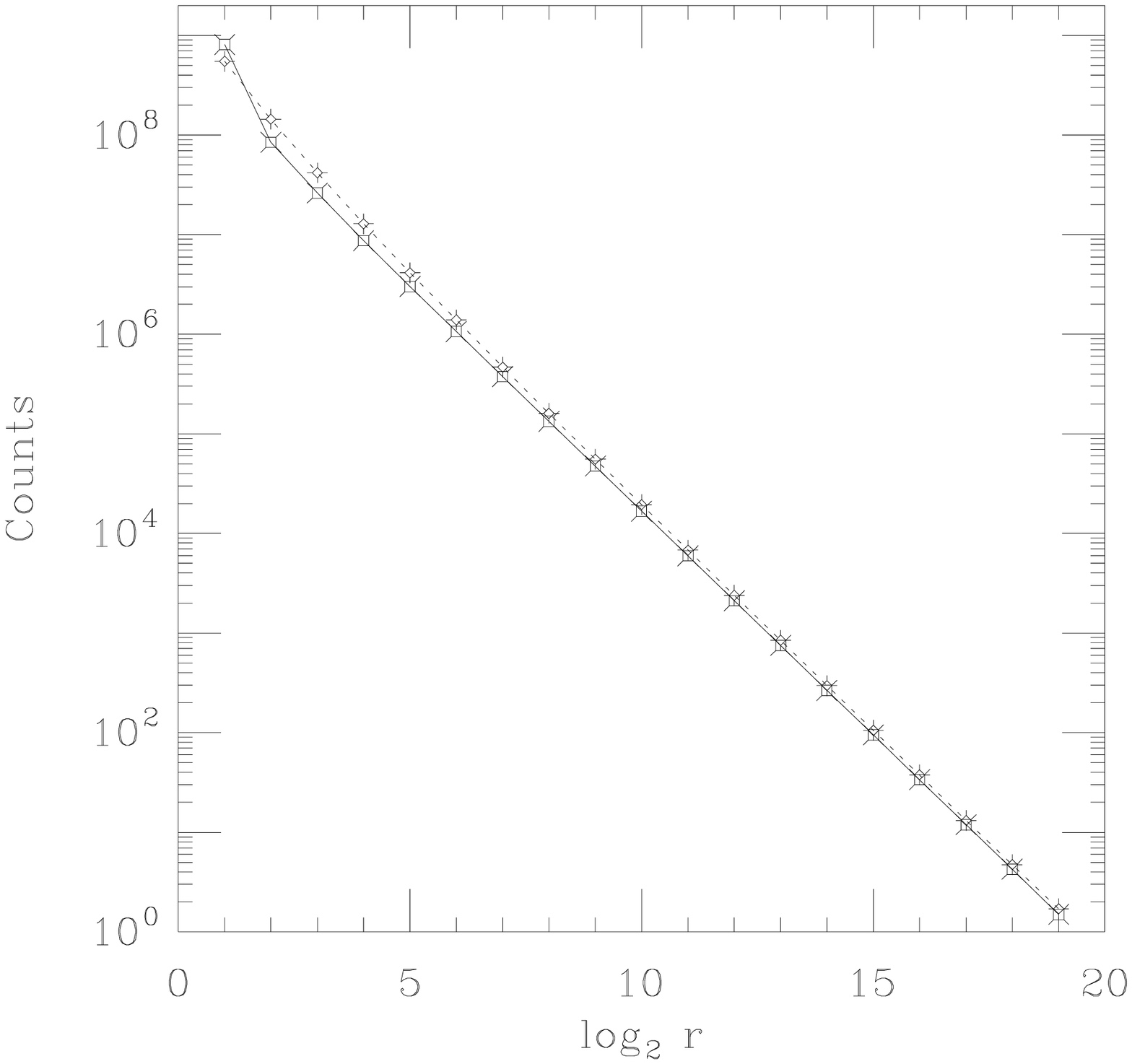}
\newpage


\begin{references}

\bibitem{soc}
P. Bak, C. Tang, and K. Wiesenfeld, Phys. Rev. Lett. {\bf 59}, 381
(1987); Phys. Rev. A. {\bf 38}, 364 (1988).
\bibitem{scaling}
M. Paczuski, S. Maslov, and P. Bak, 
Phys. Rev. E {\bf 53}, 414 (1996);
Phys. Rev. Lett. {\bf 74}, 4253 (1995);
Europhys. Lett. {\bf 27}, 97
(1994); Europhys. Lett. {\bf 28}, 295 (1994);
S. Maslov, M. Paczuski, and P. Bak, Phys. Rev. Lett. {\bf 73}, 2162
(1994). 
\bibitem{complexity}
P. Bak and M. Paczuski, Proc. Nat. Acad. Sci.  {\bf 92}, 6689 (1995).
\bibitem{G+R}
B. Gutenberg and C. F. Richter, Ann. di Geofis. {\bf 9}, 1 (1956).
\bibitem{earthquake}
P. Bak and C. Tang, J. Geophys. Res. B {\bf 94}, 15635 (1989);
Z. Olami, H.J.S. Feder, and K. Christensen, Phys. Rev. Lett.
{\bf 68},
1244 (1992);
K. Christensen and Z. Olami, Phys. Rev. A {\bf 46}, 1829 (1992);
J. Carlson and J. Langer, Phys. Rev. Lett. {\bf 62}, 2632 (1989).
\bibitem{barton}
{\it Fractals in the Earth Sciences,} eds. C.C. Barton and
P.R. Lapointe (Plenum Press, New York, 1994).
\bibitem{Gould}
S. J. Gould and N. Eldredge, Paleobiology {\bf 3}, 114 (1977); Nature
{\bf 366}, 223 (1993).
\bibitem{sfbj}
K. Sneppen, P. Bak, H. Flyvbjerg, and M.H. Jensen,
Proc. Nat. Acad. Sci. {\bf 92},
5209 (1995).
\bibitem{BS} 
P. Bak and K. Sneppen, Phys. Rev. Lett. {\bf 71}, 4083 (1993).
\bibitem{uniform}
P. Bak and M. Paczuski, in {\it Physics of Biological Systems}
  Lecture Notes in Physics (Springer-Verlag, Heidelberg , 1996).
\bibitem{ito}
K. Ito, Phys.~Rev.~E {\bf 52}, 3232 (1995).
\bibitem{lowen}
S.B. Lowen and M.C. Teich, Phys. Rev. E {\bf 47}, 992
(1993).
\bibitem{ip}
D. Wilkinson and J.F. Willemsen, J. Phys. A {\bf 16}, 3365 (1983).

\bibitem{sneppen}
K. Sneppen, Phys. Rev. Lett.  {\bf 69}, 3539 (1992).
\bibitem{zaitsev}
S.I. Zaitsev, Physica A {\bf 189}, 411 (1992).

\bibitem{otherbsmodels}
M.E.J. Newman and B.W. Roberts, Proc. Roy. Soc. Lond B {\bf 260}, 31 (1995);
K. Schmultzi H.G. Schuster, Phys. Rev. E {\bf 52}, 5273 (1995);
N. Vandewalle and M. Ausloos, J. de Phys. (Paris) {\bf 5}, 1011 (1995).
\bibitem{BoPa}
S. Boettcher and M. Paczuski, Phys. Rev. Lett. {\bf 76}, 348 (1996).
\bibitem{meanfield}
H. Flyvbjerg, K. Sneppen, and P. Bak, Phys. Rev. Lett. {\bf 71}, 4087
(1993), J. de Boer, B. Derrida, H. Flyvbjerg, A. D. Jackson, and T.
Wettig,  Phys. Rev. Lett. {\bf 73}, 906 (1994), J. de Boer, A. D.
Jackson, and T. Wettig, Phys. Rev. E {\bf 51}, 1059 (1995).
\bibitem{Kauffman}
S.A. Kauffman, {\it Origins of Order: Self-Organization and Selection
in Evolution} (Oxford University Press, Oxford 1992).
\bibitem{ultram}
R. Rammal, G. Toulouse, and M. A. Virasoro, Rev. Mod. Phys. {\bf 58},
765 (1986).
\bibitem{preprint}
M. Paczuski and S. Boettcher, preprint 1996.

\bibitem{Darwin}
C. Darwin, {\sl The Origin of Species by Means of Natural Selection, 6th
ed.,} (John Murray, London, 1910).

\bibitem{else}
S. Boettcher and M. Paczuski, in preparation.

\bibitem{sigma}
The origin of the factor preceding the Laplacian is not trivial. This
term in Eq.~(\ref{nonlocaleq}) reads in more detail ${1\over 4} \int_0^s
ds' \sigma(s-s') \nabla_r^2 F(r,s')$ with $\sigma(s)=P(s) + 2\delta(s)$
as a nonlocal diffusion constant. For large $s$ the effect of $\sigma$
becomes local, leading to the given equation.
\bibitem{BeOr}
C. M. Bender and S. A. Orszag, {\sl Advanced Mathematical Methods for
Scientist and Engineers}, (McGraw-Hill, New York, 1978).
\bibitem{letterfig}
See Fig.~1 in Ref.~\cite{BoPa}.
\bibitem{roux}
S. Roux and E. Guyon, J. Phys. A {\bf 22}, 3693 (1989).
\bibitem{maslov} 
S. Maslov, Phys. Rev. Lett. {\bf 74}, 562 (1995).
\bibitem{halpin}
T. Halpin-Healy and Y.-C. Zhang, Phys. Rep. {\bf 254}, 215 (1995).
\bibitem{tangbak}
C. Tang and P. Bak, Phys. Rev.Lett. {\bf 60}, 2347
(1988).
\bibitem{stein}
See also D.L. Stein and C.M. Newman, Phys. Rev. E {\bf 51}, 5228 (1995).
\bibitem{PSAA}
R. G. Palmer, D. L. Stein, E. Abrahams, and P. W. Anderson,
Phys.~Rev.~Lett. {\bf 53}, 958 (1984).
\bibitem{A+S}
{\sl Handbook of Mathematical Functions,} eds. M. Abramowitz and I. S.
Stegun, (Dover, New York, 1972), p. 255.

\end{references}
\end{document}